\documentclass[a4paper,12pt]{article}

\usepackage[sumlimits,intlimits,namelimits]{amsmath} 

\usepackage[english]{babel}

\usepackage{slashed}
\usepackage{phystex}

\usepackage{graphicx}

\usepackage{geometry}
\numberwithin{equation}{section}
\begin{document}
\begin{titlepage}
\begin{flushright}
SI-HEP-2006-15 \\[0.2cm]
November 13, 2006
\end{flushright}

\vspace{1.2cm}
\begin{center}
{\Large\bf
Inclusive Semi-leptonic \boldmath $B$ \unboldmath Decays \\[2mm] 
to order \ \boldmath $1/m_b^4$ \unboldmath }
\end{center}

\vspace{0.5cm}
\begin{center}
{\sc Benjamin M. Dassinger,  Thomas Mannel, Sascha Turczyk} \\[0.1cm]
{\sf Theoretische Physik 1, Fachbereich Physik,
Universit\"at Siegen\\ D-57068 Siegen, Germany}
\end{center}

\vspace{0.8cm}
\begin{abstract}
\vspace{0.2cm}\noindent
We give a systematic way to compute higher orders in the $1/m_b$ expansion 
in inclusive semi-leptonic decays at tree level. We reproduce the known $1/m_b^3$
terms and compute the $1/m_b^4$ terms at tree level. The appearing non-perturbative 
parameters and the impact on the determination of $V_{cb}$ are discussed.   
\end{abstract}

\end{titlepage}

\newpage
\pagenumbering{arabic}
\section{Introduction}
Inclusive semi-leptonic decays are the cleanest way to access the matrix 
elements of  the CKM matrix. In order to achieve precision in the determination 
of the CKM parameters a reliable theoretical framework is needed in addition 
to precise data. From the off-diagonal CKM matrix elements only $V_{us}$ and 
$V_{cb}$ are known very precisely from direct measurements \cite{PDG}, and both 
determinations can be related to a clean theoretical treatment:  
while the theoretical machinery for  the determination of $V_{us}$
from $K \to \pi \ell \bar{\nu}_\ell$ decays is chiral perturbation theory, the theoretical 
basis for the precision determination of $V_{cb}$ is the Heavy Quark Expansion 
(HQE). 

The HQE \cite{HQE0,HQE1,HQE2,HQE3}  
has become a very reliable theoretical tool, in particular for 
semi-leptonic decays. At present, 
data on inclusive semi-leptonic decays $B \to X \ell \bar{\nu}_\ell$ are 
so precise that not only a determination of $V_{cb}$ at the level of 2\% accuracy 
is possible \cite{U1,U2,S1,S2}, but also a check of the consistency 
of the HQE can be performed 
by determining the parameters of the HQE in 
different ways \cite{BF}. These parameters 
are the kinetic energy parameter $\mu_\pi$ and the chromo-magnetic moment  
$\mu_G$ at order $1/m_b^2$ and the Darwin term $\rho_D$ and the 
spin-orbit term $\rho_{\rm LS}$ at order $1/m_b^3$. We shall stick here with
the kinetic scheme, the alternative $1S$ scheme \cite{S1,S2} yields similar precision.
Using the moments of the hadronic invariant mass spectrum and the charged lepton energy spectrum these 
parameters can be consistently determined with an accuracy of about ten percent. 

Within the HQE the order of the moments is related to the order in the $1/m_b$ 
expansion \cite{FL1,FL2}. 
For example, the moments $\langle (m_b - 2 E_\ell)^n \rangle$ for the 
case of charmless semi-leptonic $B$ decays are determined by the contributions 
of the order $1/m_b^n$. Thus in order to exploit precise measurements of 
the lepton energy spectrum, i.e.\ measurements of higher moments, it is mandatory 
to  perform the theoretical calculation up to a sufficient order in the $1/m_b$ 
expansion. 

The current theoretical state of the art which is used in the fits is already quite 
elaborate. At leading order (which is the partonic rate) the full ${\cal O}(\alpha_s)$ 
and the partial  ${\cal O}(\alpha_s^2)$ result is known, while the $1/m_b^2$ and
$1/m_b^3$ results are used at tree level. Hence the leading uncertainties are 
the ${\cal O}(\alpha_s)$ contributions at $1/m_b^2$ and the tree level contributions 
to order $1/m_b^4$.  

In the present paper we present a systematic way to perform the calculation of higher 
order terms in the $1/m_b$ expansion. This approach is then used to perform  
the complete calculation of the $1/m_b^4$ terms
for the semi-leptonic decay $B \to X \ell \bar{\nu}_\ell$, keeping the charm mass to 
all orders. In the next section we shall outline our calculational method, which amounts 
to a systematization for the tree-level calculation of terms to some order $1/m_b^n$. 
This method is then applied to the case of $1/m_b^4$. In section~\ref{HQEparameters} 
we identify and discuss the parameters appearing at  $1/m_b^4$, where we find in 
total five independent matrix elements, which, however, have a simple physical   
interpretation. In section~\ref{SemiLep} we discuss the impact of these additional 
terms on the analysis of semi-leptonic decays.
 
\section{Tree-Level Operator Product Expansion to \boldmath ${\cal O} (1/m^n)$ \unboldmath} \label{OPE}
In this section we outline a method which allows us - at least in principle - to calculate
the decay rates of  semi-leptonic inclusive decays  at any order in the 
$1/m_b$ expansion at tree level. 
The starting point for such a calculation is the 
hadronic tensor, which according to the  optical theorem can be related
to the discontinuity of a time-ordered product of currents across a cut. 
Thus one starts with a correlator of two hadronic currents 
\begin{equation} \label{Tprod}
	T_{\mu \nu} = \int \text{d}^4 x e^{-i x(m_b v - q)} \bra{B(p)} \overline{b}_v(x) \Gamma_\mu c(x) \overline{c}(0)\Gamma^\dagger_\nu b_v(0) \ket{B(p)} \, .
\end{equation}
Here 
\begin{equation}  \label{sm_current}
	\Gamma_\mu = \frac{1}{2} \gamma_\mu(1-\gamma_5) 
\end{equation} 
is the left-handed current, $v = \frac{p}{M_B}$ the four velocity of the decaying B meson and $q$ the momentum transfer to the leptons. 

It is convenient to decompose this correlator into Lorentz scalar structure functions according to
\begin{equation} \label{sTprod}
	T_{\mu\nu} = -g_{\mu\nu}T_1 + v_\mu v_\nu T_2 - i \epsilon_{\mu\nu \alpha \beta} v^\alpha q^\beta T_3 + q_\mu q_\nu T_4 + \left(q_\mu v_\nu + v_\mu q_\nu \right) T_5
\end{equation}
where the scalar $T_i$ are only functions of $q^2$ and $vq$. 

Using the optical theorem we obtain for the relevant imaginary parts a 
similar decomposition:
\begin{equation}
	W_{\mu\nu} = -g_{\mu\nu}W_1 + v_\mu v_\nu W_2 - i \epsilon_{\mu\nu \alpha \beta} v^\alpha q^\beta W_3 + q_\mu q_\nu W_4 + \left(q_\mu v_\nu + v_\mu q_\nu \right) W_5 \, , 
\end{equation}
and the hadronic tensor, needed for the transition amplitude, can be computed 
from $T_{\mu\nu}$ by  
\begin{equation} \label{OT}
	-\frac{1}{\pi} \im T_j = W_j \qquad \qquad \text{ (left hand cut only)}
\end{equation}
The differential decay rate is then obtained by contracting the 
hadronic tensor with the leptonic tensor $L^{\mu\nu}$, and one obtains
\begin{equation}
	\text{d}\Gamma = \frac{G_F^2 |V_{cb}|^2}{4M_B} \im T_{\mu\nu} L^{\mu \nu} \text{d} \phi_\text{PS}
\end{equation}
where $\text{d} \phi_\text{PS}$ denotes the corresponding phase space element.  

Using the charged lepton energy $y = 2 \frac{E_l}{m_b}$, the leptonic invariant mass $\hat{q}^2 = q^2 / m_b^2$ and the rescaled total lepton energy 
$s = m_b v \hat{q}$ as independent variables one obtains the triple differential decay rate in terms of the 
scalar functions $W_i$ 
\begin{align}
	\frac{\text{d}^3\Gamma}{\text{d}\hat{q}^2 \text{d}s \text{d}y } &= 
	\frac{G_F^2 m_b^2|V_{cb}|^2}{2 \pi^3} \theta\left(m_b^2(y \frac{2s}{m_b} -y^2-\hat{q}^2) \right) \theta \left( \hat{q}^2 \right) \nonumber \\ 
  	& \quad \times \left( W_1( \hat{q}^2,s)    \,      \hat{q}^2  + W_2( \hat{q}^2,s) \, \left( y \frac{s}{m_b} -\frac{y^2}{2}-\frac{\hat{q}^2}{2}\right) + W_3 ( \hat{q}^2,s) \, \hat{q}^2\left(y m_b - s\right) \right)
\label{TDR}
\end{align}

\begin{figure}[hbpt]
	\centering\includegraphics{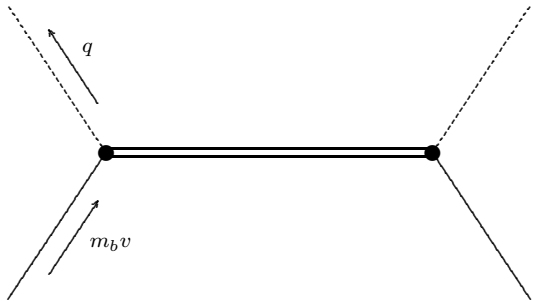}
	\caption{Tree-level Feynman diagram for the hadronic tensor in inclusive semi-leptonic $B$ decays.}
	\label{fig1}
\end{figure}

The tree-level expansion in $1/m_b$ is most easily set up by studying the Feynman  
diagram shown in fig.~\ref{fig1}. The double line denotes the propagator of a 
charm quark which is propagating in the background field of the soft gluons of the 
$B$ meson. After rescaling the $b$ quark momentum according to 
\begin{equation} \label{bquarkmom}
	p_b = m_b  v + i D 
\end{equation}
we write for the background field propagator
\begin{equation} \label{BGFprop}
	i S_{\rm BGF} = \frac{1}{\slashed{Q}+ i \slashed{D} -m_c} 
\end{equation}
where $Q=m_b v -q$ and $D$ denotes the covariant derivative with respect to the 
background gluon field. 

The tree level OPE  for semi-leptonic processes 
is obtained by multiplying (\ref{BGFprop}) 
by the appropriate Dirac matrices for the left handed current  (\ref{sm_current}).  
A calculation to order $1/m_b^n$ requires to expand this expression to $n^{th}$ 
order in the covariant derivative $(iD)$ according to 
\begin{equation} \label{BGFexpansion}
	i S_{\rm BGF} = \frac{1}{\slashed{Q}-m_c} -  \frac{1}{\slashed{Q}-m_c} ( i \slashed{D}) \frac{1}{\slashed{Q}-m_c}+ \frac{1}{\slashed{Q}-m_c} ( i \slashed{D}) \frac{1}{\slashed{Q}-m_c} ( i \slashed{D}) \frac{1}{\slashed{Q}-m_c} + \cdots 
\end{equation}
We note that this keeps track of the ordering of the covariant derivatives.  

The remaining task is to evaluate the forward matrix elements of operators of the from 
\begin{equation}
	\bar{b}_{v,\alpha} (iD_{\mu_1}) ....  (iD_{\mu_n}) b_{v,\beta}  
\end{equation} 
where $\bar{b}_v$ ($b_v$) carries the spinor indices $\alpha$ ($\beta$).  
The field $b_v$ is 
still the full QCD field, but redefined by a phase factor to remove the large piece of 
the $b$-quark momentum according to (\ref{bquarkmom}). 
This field satisfies the useful relations
\begin{align}
	b(x) &= e^{-i m_b v\cdot x} b_v(x) \label{use1} \\
	\slashed{v} b_v &=  b_v - \frac{1}{m_b} i \slashed{D} b_v \label{use2} \\
	P_+ b_v &= -\frac{1}{2m_b} i\slashed{D} b_v + b_v \label{use3} \\
	P_- b_v &= \frac{1}{2m_b} i\slashed{D} b_v \label{use4} \\
	(i v D) b_v &= -\frac{1}{2m_b}  i\slashed{D} i\slashed{D} b_v	\label{use5} 
\end{align}
which follow from the equation of motion for the $b$ quark field. Here 
$P_\pm = (1 \pm \slashed{v}) / 2$ are the projectors on the ``large'' and ``small'' 
components of a Dirac spinor. 

Note that this way of expanding will yield only 
matrix elements of local operators; however, these matrix elements contain still 
a nontrivial mass dependence, which will be discussed in the following.   

The evaluation of these matrix elements is performed  
most conveniently in a recursive fashion. The starting point is always the 
matrix element of highest dimension, i.e. the one with the maximal number of 
covariant derivatives. For these matrix elements one may neglect all contributions 
of order $1/m_b$ relative to this matrix element; in other words, this matrix element 
may be treated in the static limit. 

Thus the first step is to consider the static limit of the forward matrix element of the 
highest dimensional operator, which has the form \cite{zerec} 
\begin{align}
  	\langle B(p) | b_{v,\alpha} (iD_(\mu_1)) ....  (iD_{\mu_n}) b_{v,\beta} | B(p) \rangle  
 	&=  \langle {\rm B}_v | h_{v,\alpha} (iD_(\mu_1)) ....  (iD_{\mu_n}) h_{v,\beta} | {\rm B}_v \rangle + {\cal O} (1/m_b)  \nonumber \\   
	&=  1_{\beta \alpha} A_{\mu_1 \mu_2 \cdots \mu_n} + s_\lambda B_{\mu_1 \mu_2 \cdots \mu_n}^\lambda
\end{align}
where $s_\lambda = P_+ \gamma_\lambda \gamma_5 P_+$ is the generalization 
of the  Pauli matrices to the case $v \neq (1,0,0,0)$ and $| {\rm B}_v \rangle $ is the 
static limit of the $B$ meson state  $| B(p) \rangle$.  

The tensor structures $A$ and $B$ have to be related to a minimal set of  
fundamental matrix 
elements which are defined by contracting the indices in the various possible ways. 
In the following, 
these matrix elements  are called {\it basic parameters} for a certain order in 
$1/m_b$. For example, at order $1/m_b^3$ 
these basic parameters are  the Darwin term 
$\hat\rho_{\rm D}$ and the spin-orbit term $\hat\rho_{\rm LS}$ defined by 
\begin{align}
	2 M_B\hat\rho_{\rm D}^3 &= \langle B(p) | \bar{b}_v (i D_\mu) (i vD) (i D^\mu) b_v | B(p) \rangle \label{rhoD} \\
	2 M_B \hat\rho_{\rm LS}^3  &= \langle B(p) | \bar{b}_v (i D_\mu) (i vD) (i D_\nu) (-i \sigma^{\mu \nu})  b_v | B(p) \rangle  \label{rhoLS}
\end{align}
while at order $1/m_b^2$ the basic parameters are 
the kinetic energy parameter 
$\hat\mu_\pi$ and the chomomagnetic moment $\hat\mu_G$ given by 
\begin{align}
	2 M_B \hat\mu_\pi^2 &= -\langle B(p) | \bar{b}_v (i D)^2 b_v | B(p) \rangle \label{mupi} \\
	2 M_B \hat\mu_G^2  &= \langle B(p) | \bar{b}_v (i D_\mu) (i D_\nu) (-i \sigma^{\mu \nu} ) b_v | B(p) \rangle  \label{muG}
\end{align}

The hat above the quantitites means that we have defined these 
parameters in a covariant way. The definitions 
which have been recently used refer to the spatial components of the derivatives 
only and differ from the ones used here by higher-order terms in the $1/m$ 
expansion. The usual definitions of these quantities  are given by
\begin{align}
	2 M_B  \mu_\pi^2 &= -\langle B(p) | \bar{b}_v (i D_\perp)^2 b_v | B(p) \rangle \label{mupiu} \\
	2 M_B \hat\mu_G^2  &= \langle B(p) | \bar{b}_v (i D^\mu_\perp) (i D^\nu_\perp) (-i \sigma_{\mu \nu})  b_v | B(p) \rangle  \label{muGu} \\
	2 M_B \rho_{\rm D}^3 &= \langle B(p) | \bar{b}_v (i D_{\perp \mu}) (i vD) (i D^\mu_\perp) b_v | B(p) \rangle   \label{rhoDpiu}\\
	2 M_B \rho_{\rm LS}^3  &= \langle B(p) | \bar{b}_v (i D^\mu_\perp) (i vD) (i D^\nu_\perp) (-i \sigma_{\mu \nu})  b_v | B(p) \rangle  \label{rhoLSu} 
\end{align}
where the spatial components are  
\begin{equation}
	D^\mu_\perp =  (g^{\mu \nu} - v^\mu v^\nu) D_\nu 
\end{equation} 
At $1/m_b^4$ the correction terms in the relations between the covariant and the 
usual definition do matter and are given in the next section. 

Once the tensors $A$ and $B$ for the 
matrix elements of the highest order in the $1/m_b$ expansion have been 
expressed  in terms of these fundamental parameters, one proceeds in a similar 
way with the matrix elements of dimension $n-1$. However,  now we have to take 
into account all possible Dirac structures, such that 
\begin{equation} \label{general} 
	\langle B(p) | b_{v,\alpha} (iD_{\mu_1}) ....  (iD_{\mu_{n-1}}) b_{v,\beta} | B(p) \rangle = \sum_i \hat{\Gamma}_{\beta \alpha}^{(i)}  A_{\mu_1 \mu_2 \cdots \mu_{n-1}}^{(i)}
\end{equation}
where $\hat{\Gamma}^{(i)}$ are the complete set of the sixteen Dirac matrices. 
By using the fact that the relations (\ref{use1}-\ref{use5}) connect different orders of 
the $1/m_b$ expansion we may express the tensor coefficients $A^{(i)}$ in terms of 
the basic parameters of the order $1/m_b^{n-1}$ and the ones of the 
order $1/m_b^n$. 

This prescription defines a way to recursively compute the relevant matrix elements 
of the $1/m_b$ expansion up to order $1/m_b^n$ at tree level, starting from the 
operator of highest dimension.  Thus the leading matrix 
element of dimension three will then be expressed by this recursive method as a series 
in $1/m_b^n$ involving all the basic parameters up to this order.  

Finally, the hadronic tensor is obtained from the trace formula
\begin{align} 
	T_{\rho \sigma} &= \langle B(p) | \bar{b}_v \Gamma_\rho S_{\rm BGF}  \Gamma^\dagger_\sigma b_v | B(p) \rangle =  \sum_i  {\rm Tr}  \left\{ \Gamma_\rho  \frac{1}{\slashed{Q}-m_c}  \Gamma^\dagger_\sigma  \, \hat{\Gamma}^{(i)} \right\}   A^{(i,0)} \nonumber \\ 
	&+  \sum_i  {\rm Tr}  \left\{ \Gamma_\rho  \frac{1}{\slashed{Q}-m_c} \gamma^{\mu_1}  \frac{1}{\slashed{Q}-m_c} \Gamma^\dagger_\sigma   \, \hat{\Gamma}^{(i)}  \right\}  A^{(i,1)}_{\mu_1}  \nonumber  \\ 
	&+  \sum_i  {\rm Tr}  \left\{ \Gamma_\rho  \frac{1}{\slashed{Q}-m_c} \gamma^{\mu_1}  \frac{1}{\slashed{Q}-m_c} \gamma^{\mu_2}  \frac{1}{\slashed{Q}-m_c}\Gamma^\dagger_\sigma  \, \hat{\Gamma}^{(i)}  \right\}  A^{(i,2)}_{\mu_1 \mu_2} + \cdots   \label{Trace}
\end{align}
where the tree level expansion of the background-field propagator 
(\ref{BGFexpansion}) automatically 
yields the correct ordering of the covariant derivatives. Note that this saves us from 
computing the one- and  even more-gluon matrix elements which would be necessary 
to obtain the correct ordering of covariant derivatives in the standard computation. 

In the follwing we explicitely perform this recursion to order $1/m_b^4$, and we 
first identify the basic parameters for the order $1/m_b^4$. 

\section{Basic Parameters at \boldmath $1/m_b^4$ \unboldmath} \label{HQEparameters}
At $1/m_b^4$ we have to deal with operators of dimension 7, containing four covariant
derivatives. We find in total  five basic parameters at order $1/m_b^4$, three of which  
are spin independent, while two are spin dependent. Written in a covariant form 
we define the basic parameters of order $1/m_b^4$ to be 
\begin{align}
	2 M_B\,\, s_1 &= \bra{B(p)} \overline{b}_v i D_\rho (i v D)^2 i D^\rho b_v \ket{B(p)}  \label{p1} \\
	2 M_B\,\, s_2 &= \bra{B(p)} \overline{b}_v i D_\rho (i D)^2 i D^\rho b_v \ket{B(p)}  \\
	2 M_B\,\, s_3 &= \bra{B(p)} \overline{b}_v ((i D)^2 )^2 b_v \ket{B(p)} \\
	2 M_B\,\, s_4 &=  \bra{B(p)} \overline{b}_v i D_\mu (i D)^2 i D_\nu (-i \sigma^{\mu \nu})b_v \ket{B(p)} \\
	2 M_B\,\, s_5 &= \bra{B(p)} \overline{b}_v  i D_\rho i D_\mu i D_\nu i D^\rho(-i \sigma^{\mu \nu}) b_v \ket{B(p)} \label{p5}
\end{align}
which are real, since they are forward matrix elements of 
hermitean operators. 

In order to obtain some intuition concerning the physical meaning of these parameters, 
we may relate them to some more intuitive quantities, which are   
\begin{align}
	&\langle \vec{E}^2 \rangle : \qquad  \mbox{Expectation value of the Chromoelectric Field squared}   \\
	&\langle \vec{B}^2 \rangle : \qquad  \mbox{Expectation value of the Chromomagnetic Field squared}   \\
	&\langle (\vec{p}^2)^2 \rangle : \qquad  \mbox{Fourth power of the residual $b$ quark momentum}   \\
	&\langle (\vec{p}^2) (\vec{\sigma} \cdot \vec{B} ) \rangle : \, \, \, \mbox{Mixed Chromomagnetic Moment and res.\ Momentum squared}   \\
	&\langle (\vec{p} \cdot \vec{B}) (\vec{\sigma} \cdot \vec{p})  \rangle : \qquad \mbox{Mixed Chromomagnetic field and res.\ helicity} 
\end{align}

The basic parameters defined in (\ref{p1}-\ref{p5}) can be related to the intuitive 
quantities by\footnote{The relations between the $s_i$ and the intuitive quantities are not entirely unique. For example, $\langle (\vec{p}^2)^2 \rangle$ could also be defined by the completely symmetrized combination of the covariant derivatives}
\begin{align}
	2M_B s_1 &= -g^2 \langle \vec{E}^2 \rangle  \label{Efield} \\
	2M_B s_2 &= g^2( \langle\vec{E}^2\rangle- \langle\vec{B}^2\rangle ) + \langle\left((\vec{p})^2\right)^2\rangle \\
	2M_B s_3 &= \langle\left((\vec{p})^2\right)^2\rangle \\
	2M_B s_4 &= -3g \langle(\vec{S}\cdot\vec{B})(\vec{p})^2 \rangle + 2g\langle(\vec{p}\cdot\vec{B})(\vec{S}\cdot\vec{B})\rangle\\
	2M_B s_5 &= -g \langle(\vec{S}\cdot\vec{B})(\vec{p})^2\rangle
\end{align}
which also gives some information on the sign of the non-perturbative parameters. 

These parameters have to be determined independently from the lower-order 
parameters; however, for a numerical estimate it is useful to note that some 
of these parameters can be estimated in naive factorization, such as  
\begin{alignat}{3}
	\langle \vec{B}^2 &\rangle &\, &\sim (\langle (\vec{S}\cdot\vec{B})  \rangle)^2 &\, &=  \mu_G^4 \\ 
	\langle (\vec{p}^2)^2 &\rangle &\, &\sim   (\langle (\vec{p}^2)^2 \rangle)^2 &\, &= \mu_\pi^4 \\
	\langle (\vec{p}^2) (\vec{\sigma} \cdot \vec{B} &) &\, &\sim \langle (\vec{p}^2)^2 \rangle \langle (\vec{S}\cdot\vec{B})  \rangle &\, &= \mu_G^2 \mu_\pi^2 
\end{alignat}
while the two remaining matrix elements do not have a simple interpretation in 
naive factorization. Based on the 
interpretation of the parameters $s_i$ in terms of physical quantities we 
define a ``guestimate'' for the basic parameters $s_i$ by  
\begin{equation} \label{guess} 
	s_1 \sim - \frac{\rho_D^6}{\mu_\pi^2} \, , \quad s_2 \sim \frac{\rho_D^6}{\mu_\pi^2} - \mu_G^4 + \mu_\pi^4 \, , \quad s_3 \sim  \mu_\pi^4 \, , \quad s_4 \sim s_5 \sim - \mu_G^2 \mu_\pi^2
\end{equation} 
where at least the sign of the contributions should be correctly reproduced. 

In this way all the basic matrix elements up to order $1/m_b^4$ have been 
identified and the remaining task is to express any matrix element in terms of 
these basic quantities according to (\ref{general}). In  appendix~\ref{ME} we list all 
the necessary general matrix elements up to dimension seven in terms of the 
basic parameters $\hat\mu_\pi$, $\hat\mu_G$,  $\hat\rho_D$,  $\hat\rho_{\rm LS}$ 
and $s_1 ... s_5$.  

Finally, we may now consider the relation between the covariant definition of the 
basic parameters and the usual ones shown in (\ref{mupiu}-\ref{rhoLSu}). 
This relation is given by
\begin{align}
	\hat\mu_\pi^2 &=  \mu_\pi^2 + \frac{1}{8 m_b^2} [s_2 + s_3 + 4 s_5]    \\
	\hat\mu_G^2  &= \mu_G^2 - \frac{1}{m_b} [\rho_D^3 + \rho_{\rm LS}^3]                                               -  \frac{1}{4 m_b^2} [s_2 + s_3 + 4 s_5]   \\
	\hat\rho_D^3 &=   \rho_D^3 \vphantom{\frac{s_1}{m_b} } \\
	\hat\rho_{\rm LS}^3  &= \rho_{\rm LS}^3 - \frac{s_1}{m_b}  
\end{align}

\section{Results and Discussion} \label{SemiLep}

The remaining task is to evaluate the scalar components of the hadronic tensor
at tree level using (\ref{Trace}) and the formulae from the appendix. We first 
compute the correlator (\ref{Tprod}) involving the time-ordered product. 
From the scalar components shown in (\ref{sTprod}) we need only $T_1$, 
$T_2$  and $T_3$ due to current conservation of the leptonic current. The resulting 
expressions up to order $1/m_b^4$ are tedious and are given in  appendix~\ref{Ti}

Taking the imaginary part to obtain the components of $W_{\mu \nu}$ according to 
(\ref{OT}) we use the relation 
\begin{equation}
	- \frac{1}{\pi} {\rm Im} \Delta_0^{n+1}  =  \frac{(-1)^n}{n!} \delta^{(n)} (m_b^2 - m_c^2 + q^2 - 2 m_b \, vq )
\end{equation}
from which we can obtain the triple differential rate up to $1/m_b^4$ at tree level
using (\ref{TDR}).  The resulting expressions of the double differential rate, the 
single differential rate and the total rate are quite lengthy and given completely in  
appendix~\ref{Rates}.

The relevant quantities for the experimental analysis and the determination of 
$V_{cb}$ are the moments of the lepton energy spectrum and the moments of the 
hadronic invariant mass spectrum. In order to get a quantitative idea of the effect 
of the $1/m_b^4$ terms we consider the $1/m_b^4$ to the moments, normalized to 
the partonic rate at tree level. Thus we define 
\begin{align}
	\delta^{(4)} \langle M_X^n \rangle &= \frac{1}{\Gamma_0} \int dM_X \, M_X^n \int_{E_{\rm cut} } dE_\ell \, \frac{d^2 \Gamma^{(4)} }{dM_x \, dE_\ell}  \\
	\delta^{(4)} \langle E_\ell^n \rangle &= \frac{1}{\Gamma_0} \int dM_X \,\int_{E_{\rm cut} } dE_\ell \, E_\ell^n\frac{d^2 \Gamma^{(4)}}{dM_x \, dE_\ell} \\ 
	\Gamma_0 &= \frac{G_F^2 m_b^5 |V_{cb}|^2}{192 \pi^3} (1-8 \rho-12 \log (\rho ) \rho ^2 + 8 \rho ^3-\rho ^4) 
\end{align}
where $\Gamma^{(4)}$ is the contribution of order $1/m_b^4$. 

We first study the dependence of the ${\cal O}(1/m_b^4)$ contributions to the 
moments on the different non-perturbative parameters $s_i$. We write the 
moments as
\begin{align} 
	\delta^{(4)} \langle M_X^n \rangle &= \sum_{i=1}^5 m_b^n  f_i^{(n)} (E_{\rm cut}) \frac{s_i}{m_b^4}  \label{fs}\\
	\delta^{(4)} \langle E_\ell^n \rangle &= \sum_{i=1}^5 m_b^n  g_i^{(n)} (E_{\rm cut}) \frac{s_i}{m_b^4}  \label{gs} 
\end{align}
 
\begin{table}[hbpt]
	\centering
	\begin{tabular}{r|rrrr}
		$\smash{ _\text{\normalsize{i}} \backslash ^\text{\normalsize{n}}}$ & 1 & 2 & 3 & 4 \\
		\hline
		1 &  - 11.570 & - 6.314  & - 3.418  & - 1.845  \\
		2 &    2.073  &   1.076  &   0.556  &   0.288  \\
		3 &  - 5.969  & - 2.801  & - 1.320  & - 0.620  \\
		4 &  - 0.102  & - 0.126  & - 0.105  & - 0.074  \\
		5 &  - 3.377  & - 1.042  & - 0.174  &   0.089
	\end{tabular}
	\caption{Values of the coefficients $g_i^{(n)}$ with a weak dependence on $  E_{\rm cut}$. The values quoted are for $E_{\rm cut} = 0.8$ GeV.}
	\label{tabgs}
\end{table}
The contributions of order $1/m_b^4$ are strongly concentrated in the endpoint   
$E_\ell \sim (m_b^2-m_c^2)/(2 m_b)$ for the lepton energy. Hence the dependence 
on the cut-off energy of the coefficients is small and can be neglected, except 
for the  functions $f_i^{(1)}$, corresponding to the first hadronic mass moment.

\begin{table}[htpb]
	\centering
	\begin{tabular}{r|rrr}
		$\smash{ _\text{\normalsize{i}} \backslash ^\text{\normalsize{n}}}$ &  2 & 3 & 4 \\
	\hline
		1 &    6.214  & - 6.633  & - 1.322  \\
		2 &  - 1.343  &   1.026  &   0.203  \\
		3 &    2.472  &   1.358  & - 0.033  \\
		4 &  - 0.059  &   0.315  &   0.133  \\
		5 &  - 0.377  & - 0.129  &  0.019 
	\end{tabular}
	\caption{Values of the coefficients $f_i^{(n)}$ with a weak dependence on $  E_{\rm cut}$. The values quoted are for $E_{\rm cut} = 0.8$ GeV.}
	\label{tabfs}
\end{table}
In tables~\ref{tabgs} and \ref{tabfs}  we tabulate the values of the coefficients in 
(\ref{fs}) and (\ref{gs}) for those coefficients which are practically independent of 
$E_{\rm cut}$.
 
\begin{figure}[htpb]
	\centering\includegraphics{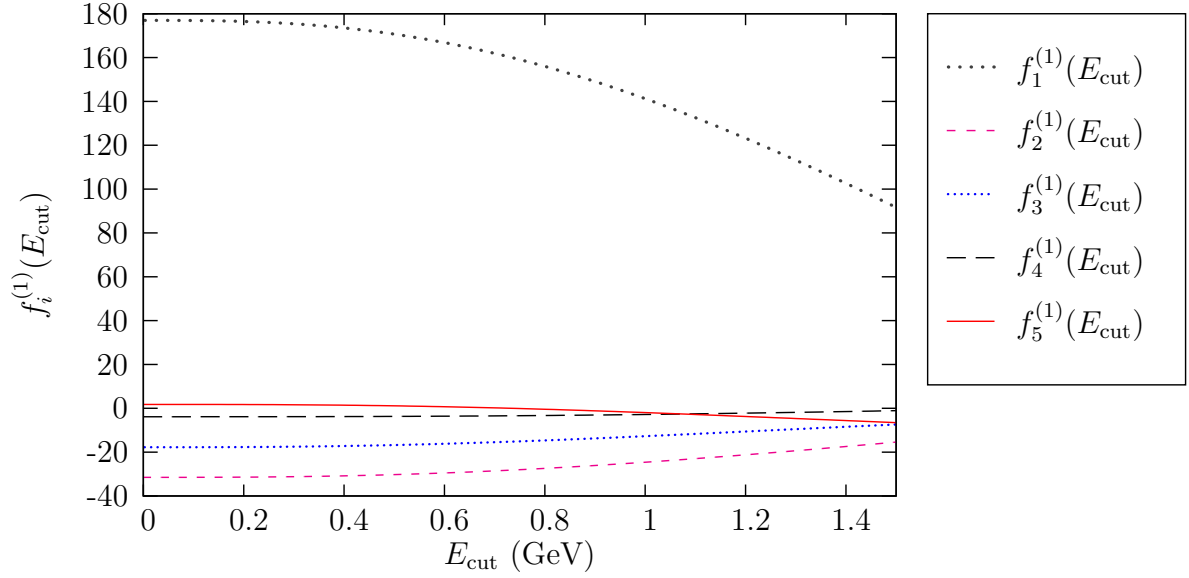}
	\caption{The dependence of $f_i^{(1)}$ on the energy cut in the lepton energy.} 
	\label{figf1a}
\end{figure} 

\begin{figure}[hbpt]
	\centering\includegraphics{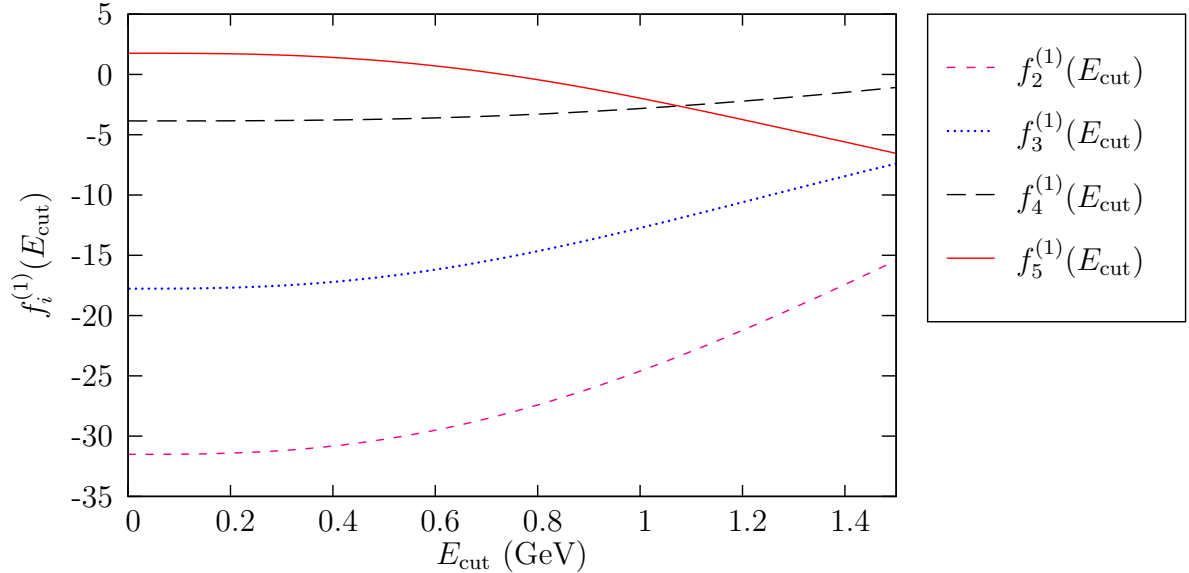}
	\caption{The dependence of $f_i^{(1)}$ on the energy cut in the lepton energy, leaving out the much larger function $f_1^{(1)}$.} 
	\label{figf1b}
\end{figure}

As pointed out above, only the coefficients $f_i^{(1)}$ of the first hadronic moment 
depend on $E_{\rm cut}$ in a substantial way. Figs.~\ref{figf1a} and \ref{figf1b}
show the dependence of the five functions on the energy cut.

Finally we shall also study the overall effect of the ${\cal O}(1/m_b^4)$ contributions. 
In order to do this we have to insert numerical values for the nonperturbative 
parameters. We use the relations (\ref{guess}) and the values obtained from the 
fit in \cite{BF} for the lower order basic parameters. In table~\ref{tabBF} we list the 
values from \cite{BF} for reference; in table~\ref{tabsi} we list the values of the 
basic parameters at $1/m_b^4$ obtained from our guestimate.

\begin{table}[htbp]
	\centering
	\begin{tabular}{c||c|c|c|c}
		Parameter & $\mu_\pi^2$ (GeV)$^2$ & $\mu_G^2$ (GeV)$^2$ & $\rho_\text{LS}^3$ (GeV)$^3$ & $\rho_D^3$ (GeV)$^3$ \\
		\hline
		Value &0.401 &0.297 &-0.183 &  0.174
	\end{tabular}
	\caption{Values taken from Buchm\"uller and Fl\"acher. We do not show the uncertainties, which are at the level of 15\%.}
	\label{tabBF}
\end{table}

\begin{table}[htbp]
	\centering
	\begin{tabular}{c||c|c|c|c|c}
		Parameter & $s_1$ (GeV)$^4$ & $s_2$ (GeV)$^4$ & $s_3$ (GeV)$^4$ & $s_4$ (GeV)$^4$ & $s_5$ (GeV)$^4$\\
		\hline
		Value &-0.076 &0.148 &0.161 & -0.119 &-0.119
	\end{tabular}
	\caption{Our guess for the basic Parameters $s_i$ which is used in the numerical analysis.}
	\label{tabsi}
\end{table}

Inserting for the masses the values $m_b = 4.59$ GeV and $m_c = 1.142$ GeV
from \cite{BF} we obtain for the contributions to the moments for a lepton energy 
cut of 0.8 GeV the  values shown in  table~\ref{tab3}. 
As pointed out above, the $1/m_b^4$ contributions to the lepton energy moments 
are practically independent of $E_{\rm cut}$, while the dependence on $E_{\rm cut}$
of the hadronic moments is given by the functions $f_i^{(1)}$. 

\begin{table}[htbp]
	\centering
	\begin{tabular}{r||r|r|r|r}
		n & 1 & 2 & 3 & 4 \\
		\hline
		$\delta^{(4)} \langle M_X^n \rangle$ & -0.1835 GeV & -0.0104 GeV$^2$ & 0.185 GeV$^3$ & 0.1064 GeV$^4$ \\
		\hline
		$\delta^{(4)} \langle E_e^n \rangle$ & 0.0066 GeV & 0.0154 GeV$^2$ & 0.0351 GeV$^3$ & 0.0803 GeV$^4$
	\end{tabular}
	\caption{Numerical values for the contribution of order $1/m_b^4$ using the parameters from table~\ref{tabsi}. }
	\label{tab3}
\end{table}

The contribution to the total rate is very small; using our estimates for the 
parameters $s_i$ we get 
\[ \frac{\delta^{(4)} \Gamma}{\Gamma} \approx 0.25\% \]
resulting in a completely negligible shift of the central value of $V_{cb}$  
from the terms of the order $1/m_b^4$.

\section{Conclusions}
In this paper we presented the complete result for the ${\cal O}(1/m_b^4)$ 
contributions to the semileptonic rate at tree level. Although we have in 
total five new, non-perturbative parameters, we can use the known matrix 
elements to order ${\cal O}(1/m_b^3)$ to estimate the size of the contributions. 

It turns out that the size of the $1/m_b^4$ terms is ``normal'', i.e. we do not 
see any abnormally large coefficients in the $1/m_b$ expansion, at least in this 
tree level calculation. 

The formulae collected in the appendix allow us to include now the 
${\cal O}(1/m_b^4)$ contributions into the moment analyses of semi-leptonic 
decays. In particular, the fourth central moment in $M_X^2$ is  dominated by the 
$1/m_b^4$ terms, so these terms are required for a sensible analysis of 
higher-order moments. 

The method suggested here allows in principle the calculation of arbitrarily high 
orders in the $1/m_b$ expansion at tree level. However, the limitation is the 
number of basic parameters, which  is presumably growing factorially, leading to 
a factorially growing effort to express the general matrix elements through the 
basic parameters.  
 
On the other hand, assuming that the $1/m_b^{\ge 5}$ terms also behave
``normally'', the uncertainty from these contributions in the determination of 
$V_{cb}$ will really be extremely small. The uncertainty currently assigned to the 
use of the HQE for the detemination of $V_{cb}$ is about 1\%; a conservative 
estimate is to use the size of the last term calculated as the uncertainty means 
that the uncertainty in the $V_{cb}$ determination due to HQE may be reduced 
to $< 1\%$.  

\subsection*{Acknowledgements}
This work was supported by the german research foundation DFG under \\
contracts no. MA1187/10-1 and KH205/1-1.

\newpage
\appendix
\section{General matrix elements in terms of basic parameters}\label{ME}
In this appendix we list the general matrix elements up to dimension seven
in terms of the parameters $\hat{\mu}_\pi$ (\ref{mupi}),   $\hat{\mu}_G$ (\ref{muG}),   
$\hat{\rho}_D$ (\ref{rhoD}),   $\hat{\rho}_{\rm LS}$ (\ref{rhoLS}), and 
$s_1 ... s_5$ (\ref{p1}-\ref{p5}).   The results are written as Dirac matrices which 
have to be inserted into the trace formula (\ref{Trace}). {\tt MATHEMATICA} 
notebooks with the corresponding expressions can be obtained from the authors.  

\subsection{Dimension seven}
\begin{alignat}{2}
 	&\langle B(p) |&\, &\bar{b}_v (iD^\rho) (iD^\sigma) (iD^\lambda) (iD^\delta) b_v | B(p)\, \rangle = \nonumber\\
	&\frac{M_B}{30}\text{P}_+  &\, &\bigg \lbrace \frac{1}{2} \bigg( (-i \sigma ^{\lambda \rho }) \left(g^{\delta \sigma }-v^{\delta } v^{\sigma }\right) + (- i \sigma ^{\delta \sigma }) \left(g^{\lambda \rho }- v^{\lambda } v^{\rho }\right)\bigg) \left(s_4+ s_5\right) \nonumber\\
	&\text{~} &\, &\quad +\bigg(- (-i \sigma ^{\rho \sigma }) \left(g^{\delta \lambda }-v^{\delta } v^{\lambda }\right) + (- i \sigma ^{\lambda \sigma }) \left(g^{\delta \rho }- v^{\delta } v^{\rho }\right)  \nonumber \\
	&\text{~} &\, &\quad + (- i \sigma ^{\delta \lambda }) \left(g^{\rho \sigma }- v^{\rho } v^{\sigma }\right)\bigg) \left(4 s_5- s_4\right) \nonumber\\
	&\text{~} &\, &\quad +(- i \sigma) ^{\delta \rho } \left(g^{\lambda \sigma }-    v^{\lambda } v^{\sigma }\right) \left(2s_4 - 3 s_5\right) \bigg \rbrace \text{P}_+ \nonumber\\ 
	+&\frac{M_B}{60} \text{P}_+  &\, &\bigg \lbrace \left(v^{\delta } v^{\lambda }- g^{\lambda \delta }\right) \left(g^{\rho \sigma }- v^{\rho } v^{\sigma }\right) \left(-2 s_1+3 s_2-7 s_3\right) \nonumber\\
	&\text{~} &\, &\quad+2 \left(g^{\delta \sigma }- v^{\delta } v^{\sigma }\right) \left(g^{\rho \lambda } - v^{\lambda } v^{\rho }\right) \left(s_1+s_2+ s_3\right) \nonumber\\
	&\text{~} &\, &\quad+\left(g^{\rho \delta }- v^{\delta } v^{\rho }\right) \left(g^{\sigma \lambda } \left(-8 s_1+7 s_2-3 s_3\right)+ v^{\lambda } v^{\sigma } \left(28 s_1-7 s_2+3 s_3\right)\right) \bigg \rbrace \nonumber\\
	&~ &\, &+{\cal O}(1/m_b) \label{MEDim7}
\end{alignat}

\subsection{Dimension six}
\begin{alignat}{2}
 	&\langle B(p) |&\, &\bar{b}_v (iD^\rho) (iD^\sigma) (iD^\lambda) b_v | B(p)\, \rangle = \nonumber \\
	&\frac{M_B}{3}\text{P}_+  &\, &\bigg \lbrace  v^{\sigma } \left(g^{\rho \lambda }- v^{\lambda } v^{\rho }\right)  \hat{\rho}_D^3 \bigg \rbrace \nonumber\\
	+&\frac{M_B}{6}  \text{P}_+ &\, &\bigg \lbrace  \left(-i \sigma ^{\lambda \rho }\right) v^{\sigma }  \hat{\rho}_\text{LS}^3 \bigg \rbrace \text{P}_+ \nonumber\\
	+&\frac{ M_B }{6 m_b} \text{P}_+  &\, &\bigg \lbrace  \left(-i \sigma ^{\lambda \rho }\right)  v^{\sigma }s_1  \bigg \rbrace\text{P}_+ \nonumber\\
	+&\frac{M_B }{12 m_b}  &\, &\bigg \lbrace \left(- i \sigma ^{\lambda \kappa}\right) v_\kappa \gamma ^{\rho } v^{\sigma }s_1  \bigg \rbrace\nonumber\\
	+&\frac{M_B}{24 m_b} &\, &\bigg \lbrace \left( (- i \sigma ^{\rho \kappa}) v_\kappa v^{\lambda } v^{\sigma }-(- i \sigma ^{\lambda \kappa}) v_\kappa v^{\rho } v^{\sigma } \right)  \left(3 s_1- s_2+s_3-s_4+s_5\right) \bigg \rbrace\nonumber\\
	+&\frac{ M_B}{48 m_b} &\, &\bigg \lbrace \left((- i \sigma ^{\rho \kappa}) v_\kappa v^{\lambda }- (i \sigma ^{\lambda \kappa}) v_\kappa v^{\rho }\right) \gamma ^{\sigma } \left(-2 s_1+s_2-s_3+2 s_5\right) \bigg \rbrace\nonumber \\
	+&\frac{M_B}{24 m_b} \text{P}_+ &\, &\bigg \lbrace   \left( (- i \sigma ^{\rho \sigma }) v^{\lambda }-(- i \sigma ^{\lambda \sigma }) v^{\rho } \right)   \left(-2 s_1+s_2-s_3+2 s_5\right)\bigg \rbrace \text{P}_+ \nonumber
\end{alignat}
\begin{alignat}{2}
	+&\frac{M_B}{24 m_b} &\, &\bigg \lbrace \left( (-i \sigma ^{\rho \sigma }) v^{\lambda }- (-i \sigma ^{\lambda \sigma }) v^{\rho } + (- i \sigma ^{\lambda \rho }) v^{\sigma }\right) \slashed{v}   \left(-s_4 +3s_5 \right) \bigg \rbrace \nonumber\\
	+&\frac{M_B}{60 m_b}   &\, &\bigg \lbrace \left(\gamma ^{\sigma }-v^{\sigma } \slashed{v} \right) \left(s_1+s_2+ s_3+s_4+s_5\right) g^{\lambda \rho } \bigg \rbrace\nonumber \\
	+&\frac{M_B}{120 m_b}  &\, &\bigg \lbrace \left(\gamma ^{\rho } \slashed{v} - v^{\rho }\right) \big((-6 \text{P}_+ -2) s_1+(4 \text{P}_+  + 3) s_2-(7-4 \text{P}_+ ) s_3 \nonumber\\
	&\text{~} &\, &\quad -10 s_5+\frac{1}{2} (8 \text{P}_+ +6) \left(s_4 + s_5\right)\big) g^{\lambda \sigma }  \bigg \rbrace \nonumber\\
	+&\frac{M_B}{12 m_b}  &\, &\bigg \lbrace \gamma ^{\lambda } \gamma ^{\rho } \slashed{v}  v^{\sigma } s_1 \bigg \rbrace\nonumber\\
	+&\frac{M_B}{120 m_b} &\, &\bigg \lbrace  \bigg(\left(\gamma ^{\lambda } \slashed{v} - v^{\lambda }\right) \left((8-6 \text{P}_+ ) s_1+(4 \text{P}_+ +3) s_3+10 s_5-\frac{1}{2} (14-8 \text{P}_+ ) \left(s_4+  s_5\right)\right) \nonumber\\
	&\text{~} &\, &\quad+(4 \text{P}_+ +3) \left(\slashed{v}  \gamma ^{\lambda }-v^{\lambda }\right) s_2\bigg) g^{\rho \sigma } \bigg \rbrace\nonumber \\
	+&\frac{ M_B}{6 m_b} &\, &\bigg \lbrace v^{\lambda } v^{\rho } v^{\sigma } \left(s_3+s_5\right) \bigg \rbrace\nonumber\\
	-&\frac{M_B}{6 m_b} \text{P}_+  &\, &\bigg \lbrace \left(v^{\rho } g^{\lambda \sigma }+ v^{\lambda } g^{\rho \sigma }\right)  \left(s_3+          s_5\right) \bigg \rbrace\nonumber\\
	-&\frac{M_B}{60 m_b} &\, &\bigg \lbrace  v^{\lambda } v^{\rho } \gamma ^{\sigma }  \left(s_1 + s_2 + s_3 + s_4 + s_5\right)\bigg \rbrace\nonumber\\
	+&\frac{M_B}{48 m_b} &\, &\bigg \lbrace \left(\gamma ^{\rho } \gamma ^{\sigma } v^{\lambda }-\gamma ^{\lambda } \gamma ^{\sigma } v^{\rho }\right) \slashed{v}   \left(-2 s_1+s_2-s_3+2 s_5\right) \bigg \rbrace \nonumber\\
	+&\frac{M_B}{120 m_b} &\, &\bigg \lbrace v^{\lambda } v^{\sigma } \gamma ^{\rho } \left(33 s_1-7 s_2+3 s_3-10 s_5-2 \left(s_4+s_5\right)\right) \bigg \rbrace\nonumber \\
	+&\frac{ M_B}{120 m_b} &\, &\bigg \lbrace v^{\rho } v^{\sigma } \gamma ^{\lambda } \left(-17 s_1+3 s_2-7 s_3+10 s_5-2 \left(s_4+s_5\right)\right) \bigg \rbrace \nonumber\\
	+&\frac{M_B }{12 m_b} &\, &\bigg \lbrace \left(\gamma ^{\rho } \gamma ^{\sigma } \gamma ^{\lambda }+\gamma^{\sigma } g^{\rho \lambda }-\gamma ^{\lambda }g^{\rho \sigma }-\gamma ^{\rho } g^{\sigma \lambda }\right) \left(\frac{1}{2} \left(s_4+ s_5\right)-2 s_5\right) \bigg \rbrace \nonumber\\
	+&\frac{M_B}{60 m_b} &\, &\bigg \lbrace v^{\lambda } v^{\rho } v^{\sigma } \slashed{v} \left(-12 s_1+3 s_2+13 s_3+ 3s_4+13 s_5 )\right) \bigg \rbrace \nonumber \\
	&~ &\, &+ {\cal O}(1/m_b^2)\label{MEDim6}
\end{alignat}

\subsection{Dimension five}
\begin{alignat}{2}
 	&\langle B(p)\, |&\, &\bar{b}_v (iD^\rho) (iD^\sigma) b_v\, |\, B(p)\, \rangle = \nonumber \\
	&\frac{ M_B }{24}\text{P}_+ &\, &\bigg \lbrace   \left(g^{\rho \sigma }- v^{\rho } v^{\sigma }\right)\left(-8 \hat{\mu}_{\pi }^2-\frac{s_2+   s_3}{m_b^2}\right) \bigg \rbrace \nonumber\\
	+&\frac{ M_B}{12}\text{P}_+  &\, &\bigg \lbrace   \left(-i \sigma ^{\rho \sigma }\right)    \left(-2 \hat{\mu}_G^2+\frac{-2s_2 -s_4 +s_5}{2m_b^2}-\frac{2 \left(\hat{\rho}_D^3+\hat{\rho} _{\text{LS}}^3\right)}{m_b}\right) \bigg \rbrace\text{P}_+ \nonumber \\
	+&\frac{M_B}{48 m_b^2} &\, &\bigg \lbrace  \bigg( (- i \sigma ^{\rho \kappa})v_\kappa v^{\sigma }-(-i \sigma ^{\sigma \kappa})v_\kappa v^{\rho }\bigg)  \left(2s_3-s_4 +9 s_5 \right) \bigg \rbrace\nonumber \\
	+&\frac{ M_B}{48 m_b^2} &\, &\bigg \lbrace (- i \sigma ^{\rho \sigma }) \left(s_2-s_3+s_4-5 s_5\right) \bigg \rbrace\nonumber \\
	+&\frac{ M_B}{6 m_b} &\, &\bigg \lbrace  \left( \text{P}_+ \gamma ^{\sigma } v^{\rho }+\gamma ^{\rho }   \text{P}_+  v^{\sigma }\right) \left(\hat{\rho}_D^3+\hat{\rho}_{\text{LS}}^3\right) \bigg \rbrace\nonumber \\
	+&\frac{M_B}{24}\text{P}_+ &\, &\bigg \lbrace   v^{\rho } v^{\sigma }  \left(\frac{3 s_2+3 s_3+16 s_5}{m_b^2}-\frac{8 \left(\hat{\rho}_D^3+ \hat{\rho}_{\text{LS}}^3\right)}{m_b}\right) \bigg \rbrace\nonumber \\
	+&\frac{M_B}{48 m_b^2}&\, &\bigg \lbrace \left(v^{\sigma } \gamma ^{\rho }+ v^{\rho } \gamma ^{\sigma }\right)  \left(4 s_1-s_2- s_3-4 s_5\right) \bigg \rbrace \nonumber\\
	+&\frac{M_B}{12 m_b^2}&\, &\bigg \lbrace \slashed{v} g^{\rho \sigma }  \left(s_1-s_2-s_4\right) \bigg \rbrace \nonumber\\
	+&\frac{M_B}{6 m_b^2}\text{P}_+ &\, &\bigg \lbrace  g^{\rho \sigma }  s_5 \bigg \rbrace\nonumber \\
	+&\frac{M_B}{24 m_b^2} &\, &\bigg \lbrace  v^{\rho } v^{\sigma } \slashed{v}  \left(-12 s_1+3 s_2 + s_3+ 2 s_4+4 s_5 \right) \bigg \rbrace \nonumber\\
	&~ &\, &+{\cal O}(1/m_b^3) \label{MEDim5}
\end{alignat}

\subsection{Dimension four}
\begin{align}
	\langle B(p) | \bar{b}_v (iD^\rho) b_v | B(p) \rangle = &-\frac{M_B}{2 m_b} P_+  \bigg\lbrace v^\rho  (\hat{\mu}^2_G-\hat{\mu}_\pi^2) \bigg \rbrace  + \frac{M_B}{6 m_b} \bigg\lbrace (\gamma^\rho -v^\rho \slashed{v}) 	    (\hat{\mu}^2_G-\hat{\mu}_\pi^2) \bigg \rbrace    \nonumber \\ 
	&+ \frac{M_B}{12 m_b^2}  \bigg\lbrace (\gamma^\rho -4\,\,v^\rho \slashed{v} ) (\hat{\rho}_\text{LS}^3 + \hat{\rho}_D^3 ) \bigg \rbrace  + {\cal O}(1/m_b^4) \label{MEDim4}
\end{align}

\subsection{Dimension 3}
\begin{equation}
   	\langle B(p) | \bar{b}_v b_v | B(p) \rangle =  P_+\,\, M_B + \frac{M_B}{4m_b^2}(\hat{\mu}^2_G-\hat{\mu}_\pi^2) + {\cal O}(1/m_b^5) \label{MEDim3}
\end{equation}
\newpage
\section{Scalar Components of the Time-Ordered Product} \label{Ti}
To compute the differential rate in the limit of vanishing lepton masses 
only the functions $T_1$, $T_2$ and $T_3$ are needed. In the following subsections 
we give the complete expressions for these scalar functions up to order $1/m^4$. 
We use the notations 
\begin{equation}
	\Delta_0 = m_b^2-2 m_b q\cdot v -m_c^2+q^2
\end{equation}
Furthermore we use dimensionless variables according to 
\begin{equation}
	\frac{m_c^2}{m_b^2} = \rho \, , \quad \frac{q^2}{m_b^2}  \rightarrow  q^2 \, , \quad m_b - v \cdot q  \rightarrow m_b v\cdot Q \, ,  \quad \frac{\Delta_0}{m_b^2} \rightarrow  \Delta_1 
 \end{equation}
{\tt MATHEMATICA} 
notebooks with the corresponding expressions can be obtained from the authors.  
 
\subsection{Scalar Function $T_1$} 
\begin{alignat}{2}
	T_1 &= \frac{v\cdot Q M_B}{m_b \Delta _1}\!\!\!\!\!&\!&\text{~} \nonumber\\
	&-\frac{M_B }{m_b^3}\hat{\mu}_\pi^2&\, &\bigg( -\frac{1}{6 \Delta _1}+\frac{ v\cdot q  (5  v\cdot q -3)-2 q^2}{3 \Delta _1^2} -\frac{4  v\cdot Q  \left((v\cdot q)^2-q^2\right)}{3 \Delta _1^3} \bigg)  \nonumber \\
	&+\frac{M_B }{3 m_b^3}\hat{\mu}_G^2 &\, &\bigg(\frac{3}{\Delta _1} + \frac{-11 q^2+10 (v\cdot q)^2+7 \rho +1}{2 \Delta _1^2} \bigg)  \nonumber \\
	&+\frac{M_B }{m_b^4}\hat{\rho}_D^3 &\, &\bigg( \frac{2}{3 \Delta _1}+\frac{4 (v\cdot q - q^2) +3 \rho +3}{3\Delta _1^2}-\frac{4 \left((v\cdot q)^2-q^2\right)  v\cdot Q }{3 \Delta _1^3} \nonumber \\
	&\text{~} &\, &+\frac{8 \left((v\cdot q)^2-q^2\right) (v\cdot Q)^2}{3 \Delta _1^4}\bigg) \nonumber \\
	&+ \frac{M_B }{m_b^4}\hat{\rho}_{LS}^3&\, &\bigg( \frac{2}{3 \Delta _1}+\frac{-4 q^2+4 (v\cdot q)^2+3 \rho -1}{3 \Delta _1^2}-\frac{4  v\cdot Q  \left((v\cdot q)^2-q^2\right)}{3 \Delta _1^3} \bigg)  \nonumber\\
	&+\frac{M_B }{m_b^5}s_1 &\, &\bigg(  \frac{2  v\cdot Q }{3 \Delta _1^2}-\frac{2 \left(-30 (v\cdot q)^3+38 (v\cdot q)^2+15 \left(q^2-4\right)  v\cdot q +7 q^2+30\right)}{15 \Delta _1^3} \nonumber \\
	&\text{~} &\, &-\frac{8 \left(q^2-(v\cdot q)^2\right) \left(-6 q^2+2  v\cdot q  (3  v\cdot q +5)+5 (\rho -2)\right)}{15 \Delta _1^4} \nonumber \\
	&\text{~} &\, &-\frac{16 \left((v\cdot q)^2-q^2\right) \left(-6 q^2+6 (v\cdot q)^2+5 \rho \right)  v\cdot Q }{15 \Delta _1^5}\bigg)   \nonumber\\
	&+ \frac{M_B }{m_b^5}s_2 &\, &\bigg( \frac{1-3  v\cdot q }{6 \Delta _1^2}+\frac{(35  v\cdot q -23) q^2-2  v\cdot q  ( v\cdot q  (25  v\cdot q -39)+40)+40}{30 \Delta _1^3} \nonumber \\
	&\text{~} &\, &-\frac{2 \left(q^2-6 (v\cdot q)^2+5 \rho +5\right) \left(q^2-(v\cdot q)^2\right)}{15 \Delta _1^4} + \frac{8  v\cdot Q  \left(q^2-(v\cdot q)^2\right)^2}{5 \Delta _1^5}\bigg)  \nonumber \\
	&+\frac{M_B }{m_b^5}s_3 &\, &\bigg(\frac{ v\cdot q +5}{6 \Delta _1^2}+\frac{3 (5  v\cdot q -1) q^2-2  v\cdot q  ( v\cdot q  (15  v\cdot q +11)-40)-40}{30 \Delta _1^3} \nonumber \\
	&\text{~} &\, &-\frac{4 \left(3 q^2+(10-13  v\cdot q )  v\cdot q \right) \left(q^2-(v\cdot q)^2\right)}{15 \Delta _1^4}+\frac{8  v\cdot Q  \left(q^2-(v\cdot q)^2\right)^2}{5 \Delta _1^5}\bigg) \nonumber
\end{alignat}
\begin{alignat}{2}
	&+\frac{M_B }{m_b^5}s_4 &\, &\bigg(-\frac{7  v\cdot q +1}{12 \Delta _1^2}+\frac{2 ( v\cdot q +1) \left(q^2-(v\cdot q)^2\right)}{3  \Delta _1^3} \nonumber \\
	&\text{~} &\, &-\frac{2 \left(q^2-(v\cdot q)^2\right) \left(q^2-2 (v\cdot q)^2+\rho +1\right)}{5 \Delta _1^4} \bigg) \nonumber  \\
	&+\frac{M_B }{m_b^5}s_5 &\,&\bigg( \frac{3  v\cdot q +5}{4 \Delta _1^2}+\frac{2 \left(q^2 (3  v\cdot q +1)-2  v\cdot q  (2-3  v\cdot Q   v\cdot q )\right)}{3 \Delta _1^3} \nonumber \\
	&\text{~} &\, &+\frac{4 \left(q^2-(v\cdot q)^2\right) \left(-19 q^2+13 (v\cdot q)^2+16 \rho +6\right)}{15 \Delta _1^4} \bigg)
\end{alignat}
\subsection{The scalar function $T_2$}
\begin{alignat}{2}
	T_2 &=  \frac{2 M_B}{m_b \Delta _1}&\! &\text{~} \nonumber\\
	&-\frac{M_B }{m_b^3}\hat{\mu}_\pi^2 &\, &\bigg(-\frac{5}{3\Delta _1}-\frac{14  v\cdot q }{3 \Delta _1^2}+\frac{8 \left(q^2-(v\cdot q)^2\right)}{3 \Delta _1^3}\bigg)  \nonumber \\
	&+\frac{M_B }{3 m_b^3}\hat{\mu}_G^2 &\, &\bigg( -\frac{5}{\Delta _1} +\frac{4-10  v\cdot q }{\Delta _1^2}\bigg)  \nonumber \\
	&+\frac{M_B }{m_b^4}\hat{\rho}_D^3 &\, &\bigg( -\frac{4}{3 \Delta _1} +\frac{4-4  v\cdot q }{ \Delta _1^2}+\frac{16  v\cdot Q   v\cdot q }{3  \Delta _1^3}-\frac{16  v\cdot Q  \left(q^2-(v\cdot q)^2\right)}{3 \Delta _1^4}\bigg) \nonumber \\
	&+\frac{M_B }{m_b^4}\hat{\rho}_{LS}^3 &\, &\bigg(-\frac{4}{3 \Delta _1}+ \frac{2 - 3  v\cdot q }{12 \Delta _1^2}-\frac{4 \left(q^2-2  v\cdot Q   v\cdot q \right)}{3 \Delta _1^3} \bigg)  \nonumber\\
	&+\frac{M_B }{m_b^5}s_1 &\, &\bigg( -\frac{4 (3  v\cdot q +5)}{3 \Delta _1^2}-\frac{4 \left(-39 q^2+42 (v\cdot q)^2+40  v\cdot q +35 \rho -5\right)}{15 \Delta _1^3} \nonumber \\
	&\text{~} &\, &-\frac{32 \left((6  v\cdot q +5) \left((v\cdot q)^2-q^2\right)+5  v\cdot q  \rho \right)}{15 \Delta _1^4} \nonumber \\
	&\text{~} &\, &+\frac{32 \left(q^2-(v\cdot q)^2\right) \left(-6 q^2+6 (v\cdot q)^2+5 \rho \right)}{15 \Delta _1^5}\bigg)   \nonumber\\
	&+\frac{M_B }{m_b^5}s_2 &\, &\bigg(\frac{5  v\cdot q +3}{3 \Delta _1^2}+\frac{-39 q^2+62 (v\cdot q)^2+5 \rho +15}{15 \Delta_1^3} \nonumber \\
	&\text{~} &\, &+\frac{8 (6  v\cdot q +5) \left((v\cdot q)^2-q^2\right)}{15 \Delta _1^4}+\frac{16 \left(q^2-(v\cdot q)^2\right)^2}{5 \Delta _1^5}\bigg)  \nonumber \\
	&+\frac{M_B }{m_b^5}s_3 &\, &\bigg( \frac{3  v\cdot q -1}{3 \Delta _1^2}+\frac{q^2+82 (v\cdot q)^2-25 \rho +5}{15 \Delta_1^3}\nonumber \\
	&\text{~} &\, &+\frac{128  v\cdot q  \left((v\cdot q)^2-q^2\right)}{15 \Delta _1^4}+\frac{16 \left(q^2-(v\cdot q)^2\right)^2}{5 \Delta _1^5}\bigg) \nonumber  \\
	&+\frac{M_B }{m_b^5}s_4 &\, &\bigg(\frac{4  v\cdot q +1}{6 \Delta _1^2}-\frac{4 \left(q^2+(5-8  v\cdot q )  v\cdot q \right)}{15 \Delta _1^3}-\frac{4  v\cdot Q  \left(q^2+2(v\cdot q)^2\right)}{5 \Delta _1^4} \bigg) \nonumber \\	
	&+\frac{M_B }{m_b^5}s_5 &\, &\bigg( \frac{24  v\cdot q +13}{6 \Delta _1^2}+\frac{-94 q^2+172 (v\cdot q)^2-20 \rho +20}{15 \Delta _1^3}\nonumber \\
	&\text{~} &\, &+\frac{4 \left((17-47  v\cdot q ) q^2+2 (v\cdot q)^2 (13  v\cdot q +2)\right)}{15 \Delta _1^4}\bigg)
\end{alignat}

\subsection{The scalar function $T_3$}
\begin{alignat}{2}
	T_3 &=  \frac{M_B}{m_b^2 \Delta _1} &\!	&\text{~} \nonumber\\
	&-\frac{M_B }{m_b^4}\hat{\mu}_\pi^2 &\, &\bigg(-\frac{5  v\cdot q }{3 \Delta _1^2}+\frac{4 \left(q^2-(v\cdot q)^2\right)}{3 \Delta _1^3}  \bigg)  \nonumber \\
	&+\frac{M_B }{3 m_b^4}\hat{\mu}_G^2 &\, &\bigg( \frac{6-5  v\cdot q }{\Delta _1^2} \bigg)  \nonumber \\
	&+\frac{M_B }{m_b^5}\hat{\rho}_D^3 &\, &\bigg(\frac{6-4  v\cdot q }{3 \Delta _1^2}+\frac{4  v\cdot Q   v\cdot q }{3 \Delta _1^3}-\frac{8  v\cdot Q  \left(q^2-(v\cdot q)^2\right)}{3 \Delta _1^4} \bigg) \nonumber \\
	&+\frac{M_B }{m_b^5}\hat{\rho}_{LS}^3 &\, &\bigg( \frac{6-4  v\cdot q }{3 \Delta _1^2}-\frac{4 (v\cdot Q)^2}{3 \Delta _1^3}\bigg)  \nonumber\\
	&+\frac{M_B }{m_b^6}s_1 &\, &\bigg( \frac{2 \left(q^2-6 (v\cdot q)^2+8  v\cdot q -6\right)}{3 \Delta _1^3}+\frac{16 (3  v\cdot q +5) \left(q^2-(v\cdot q)^2\right)-40  v\cdot q  \rho }{15 \Delta _1^4} \nonumber \\
	 &\text{~} &\, &+\frac{16 \left(q^2-(v\cdot q)^2\right) \left(-6 q^2+6 (v\cdot q)^2+5 \rho \right)}{15 \Delta _1^5} \bigg)   \nonumber\\
	&+\frac{M_B }{m_b^6} s_2&\, &\bigg( \frac{5}{3 \Delta _1^2}+\frac{-9 q^2+10 (v\cdot q)^2+4 \rho +2}{6 \Delta _1^3} \nonumber \\
	&\text{~} &\, &+\frac{4(3  v\cdot q +5) \left((v\cdot q)^2-q^2\right)}{15 \Delta _1^4}+\frac{8 \left(q^2-(v\cdot q)^2\right)^2}{5 \Delta _1^5}\bigg)  \nonumber \\
	&+\frac{M_B }{m_b^6}s_3 &\, &\bigg( -\frac{1}{\Delta _1^2}+\frac{5 q^2+6((v\cdot q)^2-\rho) }{6 \Delta _1^3}+\frac{52  v\cdot q  \left((v\cdot q)^2-q^2\right)}{15\Delta _1^4}+\frac{8 \left(q^2-(v\cdot q)^2\right)^2}{5 \Delta _1^5}\bigg) \nonumber \\
	&+\frac{M_B }{m_b^6}s_4 &\, &\bigg(\frac{1}{4 \Delta _1^2}-\frac{2 \left(\rho -(v\cdot q)^2\right)}{3 \Delta _1^3}+ \frac{4  v\cdot q  \left((v\cdot q)^2-q^2\right)}{5 \Delta _1^4} \bigg) \nonumber  \\
	&+\frac{M_B }{m_b^6}s_5 &\, &\bigg( \frac{35}{12 \Delta _1^2}+\frac{-11 q^2+12 (v\cdot q)^2+9 \rho -9}{3 \Delta _1^3}+\frac{4 (13  v\cdot q -25) \left((v\cdot q)^2-q^2\right)}{15 \Delta _1^4}\bigg)
\end{alignat}

\newpage
\section{Decay Rates} \label{Rates}
Finally we list in this appendix the full expressions for the differential and the total 
decay rate at tree level up to order $1/m_b^4$. {\tt MATHEMATICA} 
notebooks with the corresponding expressions can be obtained from the authors.  

\subsection{Double Differential rate $\text{\normalfont{d}}^2 \Gamma/(\text{\normalfont{d}}\hat{E}_0 \, \text{\normalfont{d}}M_X^2)$}
We use for this double differential rate the variables
\begin{equation}
	E_0  = m_b - v\cdot q \, , \quad M_X^2 = (m_b v - q)^2
\end{equation}
\begin{alignat}{2}
	\frac{\text{d} \Gamma^{(2)}}{\text{d} \hat{M}^2_X \hat{E}_0} = &\frac{G_F^2 m_b^5}{192 \pi^3} |V_{cb}|^2 &\, &\sqrt{\hat{E}_0^2-\hat{M}_X^2}\bigg\lbrace \bigg(16  \left(4 \hat{E}_0^2-3 \left(\hat{M}_X^2+1\right) \hat{E}_0+2 \hat{M}_X^2\right) \bigg) \delta(\hat{M}^2_X - \rho)  \nonumber \\
	 &+ \frac{8 }{3 m_b^2}\hat{\mu}_\pi^2 &\, &\bigg( \bigg(40 \hat{E}_0^3-  2 \hat{E}_0^2 \left(10 \hat{M}_X^2 + 5 \rho + 11\right) - 3 \rho  \nonumber \\
	&~ &\, &+ \hat{E}_0 \left(6 \rho - 10 \hat{M}_X^2\right)+5 \hat{M}_X^4 + 7 (\rho + 1) \hat{M}_X^2\bigg) \delta'(\hat{M}^2_X - \rho)  \nonumber \\
	&~ &\, &+\bigg(16 \hat{E}_0^4 - 4 \left(2 \hat{M}_X^2 + \rho + 3\right) \hat{E}_0^3 - 8 \rho  \hat{E}_0^2+ 2 \hat{M}_X^2 \left(\hat{M}_X^2 - 5 \rho \right) \nonumber \\
	&~ &\, &+ \left(-\hat{M}_X^4 + (13 \rho + 3) \hat{M}_X^2 + 9 \rho \right) \hat{E}_0 \bigg)\delta''(\hat{M}^2_X -\rho)\bigg) \nonumber \\
	&+ \frac{8}{3 m_b^2}\hat{\mu}_G^2 &\, &\bigg(-\bigg(  + 40 \hat{E}_0^3- 2 \hat{E}_0^2 \left(10 \hat{M}_X^2 + 5 \rho + 11\right) \nonumber \\
	&~ &\, &+ \hat{E}_0 \left(6 \rho - 10 \hat{M}_X^2\right)  + 5 \hat{M}_X^4 + 7 (\rho + 1) \hat{M}_X^2 - 3 \rho \bigg)\delta'(\hat{M}^2_X -\rho)  \nonumber \\
	&~ &\, &+\bigg(  \left(8 \hat{E}_0^2 - 3 \left(\hat{M}_X^2 + 1\right) \hat{E}_0 -  2 \hat{M}_X^2\right)\left(\hat{M}_X^2 - \rho \right)\bigg)\delta''(\hat{M}^2_X -\rho)\bigg) \nonumber \\
	&+ \frac{8}{3 m_b^3} \hat{\rho}_D^3&\, &\bigg(-2 \bigg( 16 \hat{E}_0^3- 2 \hat{E}_0^2 \left(4 \hat{M}_X^2 + 2 \rho + 5\right)   + \hat{E}_0 \left(6 \rho - 4 \hat{M}_X^2\right) \nonumber \\
	&~ &\, &+2 \hat{M}_X^4 + (\rho + 4) \hat{M}_X^2 - 3 \rho \bigg) \delta'(\hat{M}^2_X -\rho)  \nonumber \\
	&~ &\, &-4\left(\hat{E}_0^2 - \hat{M}_X^2\right) \left(4 \hat{E}_0^2 - \left(2 \hat{M}_X^2 + \rho + 1\right) \hat{E}_0 - \hat{M}_X^2 + \rho \right) \delta''(\hat{M}^2_X -\rho) \nonumber \\
 	&~ &\, &-\frac{1}{3} \bigg(32 \hat{E}_0^5 - 8 \left(2 \hat{M}_X^2 + \rho + 3\right) \hat{E}_0^4 + 8 \left(\rho - 3 \hat{M}_X^2\right) \hat{E}_0^3  \nonumber \\
	&~ &\, &+ 2 \left(5 \hat{M}_X^4 + 2 (5 \rho + 6) \hat{M}_X^2 - 3 \rho ^2\right) \hat{E}_0^2 + 2 \left(5 \hat{M}_X^4 - 22 \rho  \hat{M}_X^2 + 9 \rho ^2\right) \hat{E}_0  \nonumber \\
	&~ &\, &- 3 \left(\hat{M}_X^2 + 3\right) \left(\hat{M}_X^2 - \rho \right)^2\bigg)\delta^{(3)}(\hat{M}^2_X -\rho) \bigg) \nonumber \\
	&+ \frac{8}{3 m_b^3}\hat{\rho}_{LS}^3 &\, &\bigg(-2\bigg(+ 16 \hat{E}_0^3 - 2 \hat{E}_0^2 \left(4 \hat{M}_X^2 + 2 \rho + 5\right)  + \hat{E}_0 \left(6 \rho - 4 \hat{M}_X^2\right) \nonumber \\
	&~ &\, &+2 \hat{M}_X^4 + (\rho + 4) \hat{M}_X^2- 3 \rho \bigg)\delta'(\hat{M}^2_X -\rho)  \nonumber 
\end{alignat}
\begin{alignat}{2}
	&~ &\, &- 4\left(\hat{E}_0^2 - \hat{M}_X^2\right) \left(4 \hat{E}_0^2 - \left(2 \hat{M}_X^2 + \rho + 1\right) \hat{E}_0 - \hat{M}_X^2 + \rho \right)\delta''(\hat{M}^2_X -\rho) \nonumber \\
 	&~ &\, &+\frac{1}{3}\bigg(8 \hat{E}_0^3 \left(\hat{M}_X^2 - \rho \right) - 2 \hat{E}_0^2 \left(\hat{M}_X^4 + 2 \hat{M}_X^2 - \rho  (\rho + 2)\right) \nonumber \\
	&~ &\, & - 2 \hat{E}_0 \left(\hat{M}_X^4 + 2 \rho  \hat{M}_X^2 - 3 \rho ^2\right) -\hat{M}_X^6 + (6 \rho + 1) \hat{M}_X^4 \nonumber \\
	 &~ &\, &+ (2 - 5 \rho ) \rho  \hat{M}_X^2 - 3 \rho ^2\bigg)\delta^{(3)}(\hat{M}^2_X -\rho) \bigg) \nonumber \\
	&+ \frac{4}{3 m_b^4} s_1 &\, &\bigg(-4 \bigg(  24 \hat{E}_0^4 - 2 \left(6 \hat{M}_X^2 + 3 \rho + 7\right) \hat{E}_0^3 + \left(8 \rho - 12 \hat{M}_X^2\right) \hat{E}_0^2 \nonumber \\
	 &~ &\, &+ \left(6 \hat{M}_X^4 + (3 \rho + 8) \hat{M}_X^2 - 3 \rho \right) \hat{E}_0 - 2 \rho  \hat{M}_X^2\bigg)\delta''(\hat{M}^2_X -\rho) \nonumber \\
 	&~ &\, & -\frac{16}{15} \left(\hat{E}_0^2 - \hat{M}_X^2\right) \bigg(24 \hat{E}_0^3 - 6 \left(2 \hat{M}_X^2 + \rho + 1\right) \hat{E}_0^2  \nonumber \\
	&~ &\, &+ \left(5 \rho - 9 \hat{M}_X^2\right) \hat{E}_0 + \hat{M}_X^2 \left(2 \hat{M}_X^2 + \rho + 1\right)\bigg)\delta^{(3)}(\hat{M}^2_X -\rho) \nonumber \\
	&~ &\, &-\frac{1}{15} \bigg( 192 \hat{E}_0^6 - 48 \left(2 \hat{M}_X^2 + \rho + 3\right) \hat{E}_0^5 + \left(48 \rho - 176 \hat{M}_X^2\right) \hat{E}_0^4  \nonumber \\
	&~ &\, &+ 8 \left(9 \hat{M}_X^4 + (17 \rho + 21) \hat{M}_X^2 - 5 \rho ^2\right) \hat{E}_0^3+ 4 \left(19 \hat{M}_X^4 - 64 \rho  \hat{M}_X^2 + 25 \rho ^2\right) \hat{E}_0^2 \nonumber \\
	&~ &\, & - \left(21 \hat{M}_X^6 + (69 - 2 \rho ) \hat{M}_X^4 + 5 (\rho - 18) \rho  \hat{M}_X^2 + 45 \rho ^2\right) \hat{E}_0 \nonumber \\
	&~ &\, & - 2 \hat{M}_X^2 \left(\hat{M}_X^4 - 14 \rho  \hat{M}_X^2 + 5 \rho ^2\right)\bigg)\delta^{(4)}(\hat{M}^2_X -\rho)\bigg) \nonumber \\
	&+ \frac{1}{3 m_b^4}s_2 &\, &\bigg(4\bigg( 40 \hat{E}_0^4 - 2 \left(10 \hat{M}_X^2 + 5 \rho + 13\right) \hat{E}_0^3 - 4 \left(\hat{M}_X^2 + 2 \rho \right) \hat{E}_0^2 \nonumber \\
	&~ &\, &+ \left(2 \hat{M}_X^4 + (19 \rho + 8) \hat{M}_X^2 + 9 \rho \right) \hat{E}_0 - 10 \rho  \hat{M}_X^2 \bigg)\delta''(\hat{M}^2_X -\rho) \nonumber \\
 	&~ &\, &+\frac{8}{15}\bigg(   48 \hat{E}_0^5-12 \hat{E}_0^4 \left(2 \hat{M}_X^2 + \rho + 1\right) - 6 \hat{E}_0^3 \left(11 \hat{M}_X^2 + 5 \rho \right)\nonumber \\
	&~ &\, & + 2 \hat{E}_0^2 \left(14 \hat{M}_X^4 + (17 \rho + 7) \hat{M}_X^2 + 5 \rho  (\rho + 1)\right) + 6 \hat{E}_0 \left(3 \hat{M}_X^4 + 5 \rho ^2\right)\nonumber \\
	&~ &\, &   -4 \hat{M}_X^6 - (7 \rho + 2) \hat{M}_X^4 + 5 (1 - 5 \rho ) \rho  \hat{M}_X^2 - 15 \rho ^2  \bigg)\delta^{(3)}(\hat{M}^2_X -\rho) \nonumber \\
	&~ &\, &+\frac{1}{15}\bigg(  192 \hat{E}_0^6 - 48 \left(2 \hat{M}_X^2 + \rho + 3\right) \hat{E}_0^5 - 16 \left(11 \hat{M}_X^2 + 7 \rho \right) \hat{E}_0^4  \nonumber \\
	&~ &\, &+ 24 \left(3 \hat{M}_X^4 + (9 \rho + 7) \hat{M}_X^2 + 5 \rho \right) \hat{E}_0^3  + 4 \left(19 \hat{M}_X^4 - 34 \rho  \hat{M}_X^2 + 15 \rho ^2\right) \hat{E}_0^2 \nonumber \\
	&~ &\, & - 3 \left(7 \hat{M}_X^6 + (26 \rho + 23) \hat{M}_X^4 + 5 \rho  (3 \rho + 2) \hat{M}_X^2 + 15 \rho ^2\right) \hat{E}_0  \nonumber \\
	&~ &\, &- 2 \hat{M}_X^2 \left(\hat{M}_X^4 - 34 \rho  \hat{M}_X^2 - 15 \rho ^2\right)\bigg)\delta^{(4)}(\hat{M}^2_X -\rho) \bigg) \nonumber \\
	&+  \frac{1}{3 m_b^4}s_3 &\, &\bigg( 4\bigg(  24 \hat{E}_0^4 - 2 \left(6 \hat{M}_X^2 + 3 \rho + 7\right) \hat{E}_0^3 + \left(8 \rho - 12 \hat{M}_X^2\right) \hat{E}_0^2 \nonumber 
\end{alignat}
\begin{alignat}{2}
	&~ &\, &+ \left(6 \hat{M}_X^4 + (3 \rho + 8) \hat{M}_X^2 - 3 \rho \right) \hat{E}_0 - 2 \rho  \hat{M}_X^2\bigg)\delta''(\hat{M}^2_X -\rho) \nonumber \\
 	&~ &\, &+\frac{8}{15}\bigg( 208 \hat{E}_0^5- 4 \hat{E}_0^4 \left(26 \hat{M}_X^2 + 13 \rho + 33\right)  - 2 \hat{E}_0^3 \left(23 \hat{M}_X^2 + 65 \rho \right)  \nonumber \\
	&~ &\, & + 2 \hat{E}_0^2 \left(14 \hat{M}_X^4 + (67 \rho + 27) \hat{M}_X^2 + 15 \rho  (\rho + 3)\right) - 2 \hat{E}_0 \left(21 \hat{M}_X^4 - 50 \rho  \hat{M}_X^2 + 45 \rho ^2\right)\nonumber \\
	&~ &\, & + 16 \hat{M}_X^6 + (18 - 67 \rho ) \hat{M}_X^4 + 15 (\rho - 5) \rho  \hat{M}_X^2  + 45 \rho ^2\bigg)\delta^{(3)}(\hat{M}^2_X -\rho) \nonumber \\
	&~ &\, &+\frac{1}{15}\bigg(  192 \hat{E}_0^6 - 48 \left(2 \hat{M}_X^2 + \rho + 3\right) \hat{E}_0^5 - 16 \left(\hat{M}_X^2 + 17 \rho \right) \hat{E}_0^4 \nonumber \\
	&~ &\, &+ 16 \left(2 \hat{M}_X^4 + (11 \rho + 3) \hat{M}_X^2 + 5 \rho  (\rho + 3)\right) \hat{E}_0^3 - 4 \left(21 \hat{M}_X^4 - 26 \rho  \hat{M}_X^2 +  5 \rho ^2\right) \hat{E}_0^2 \nonumber \\
	&~ &\, & + \left(19 \hat{M}_X^6 + (51 - 38 \rho ) \hat{M}_X^4 - 25 \rho  (5 \rho + 6) \hat{M}_X^2 - 45 \rho ^2\right) \hat{E}_0 \nonumber \\
	&~ &\, &-  2 \hat{M}_X^2 \left(\hat{M}_X^4 + 6 \rho  \hat{M}_X^2 - 55 \rho ^2\right)\bigg)\delta^{(4)}(\hat{M}^2_X -\rho)\bigg) \nonumber \\
	&+\frac{2}{3 m_b^4}s_4 &\, &\bigg(\bigg(32 \hat{E}_0^4 - 8 \left(2 \hat{M}_X^2 + \rho + 3\right) \hat{E}_0^3 + 8 \left(\hat{M}_X^2 - 3 \rho \right) \hat{E}_0^2  \nonumber \\
	&~ &\, &+ \left(-5 \hat{M}_X^4 + (29 \rho + 3) \hat{M}_X^2 + 21 \rho \right) \hat{E}_0 + 2 \hat{M}_X^2 \left(\hat{M}_X^2 - 9 \rho \right) \bigg)\delta''(\hat{M}^2_X -\rho) \nonumber \\
 	&~ &\, &+\frac{8}{15}\bigg(  24 \hat{E}_0^5- 6 \hat{E}_0^4 \left(2 \hat{M}_X^2 + \rho + 1\right) + 2 \hat{E}_0^3 \left(\hat{M}_X^2 - 25 \rho \right)     \nonumber\\
	&~ &\, &+ \hat{E}_0^2 \left(-6 \hat{M}_X^4 + (47 \rho - 13) \hat{M}_X^2 - 5 (\rho - 5) \rho \right)+  2 \hat{E}_0 \left(2 \hat{M}_X^4 - 5 \rho  \hat{M}_X^2 + 15 \rho^2\right) \nonumber \\
	 &~ &\, & +3 \hat{M}_X^6 + (4 - 11 \rho ) \hat{M}_X^4 + 5 (1 - 2 \rho ) \rho  \hat{M}_X^2- 15 \rho ^2\bigg)\delta^{(3)}(\hat{M}^2_X -\rho) \nonumber \\
	&~ &\, & -\frac{4}{15} \left(\hat{E}_0^2 - \hat{M}_X^2\right) \left(\hat{M}_X^2 - \rho \right) \left(-6 \hat{E}_0^2 + 5 \left(\hat{M}_X^2 - \rho \right) \hat{E}_0 + \hat{M}_X^2 + 5 \rho \right)\bigg)\delta^{(4)}(\hat{M}^2_X -\rho) \nonumber \\
	&+ \frac{2}{3 m_b^4} s_5 &\, &\bigg(\bigg(  192 \hat{E}_0^4 - 16 \left(6 \hat{M}_X^2 + 3 \rho + 7\right) \hat{E}_0^3 - 8 \left(13 \hat{M}_X^2 - 9 \rho \right) \hat{E}_0^2  \nonumber \\
	&~ &\, &+ \left(51 \hat{M}_X^4 + (21 \rho + 67) \hat{M}_X^2 - 27 \rho \right) \hat{E}_0 + 2 \hat{M}_X^2 \left(\hat{M}_X^2 - 9 \rho \right)\bigg)\delta''(\hat{M}^2_X -\rho) \nonumber \\
 	&~ &\, &+\frac{8}{15}\bigg(  104 \hat{E}_0^5- 2 \hat{E}_0^4 \left(26 \hat{M}_X^2 + 13 \rho + 33\right)- 8 \hat{E}_0^3 \left(16 \hat{M}_X^2 - 5 \rho \right)  \nonumber \\
	&~ &\, &+ \hat{E}_0^2 \left(74 \hat{M}_X^4 + (87 - 23 \rho ) \hat{M}_X^2 + 15 \rho  (3 \rho - 1)\right) + \hat{E}_0 \left(-6 \hat{M}_X^4 + 80 \rho  \hat{M}_X^2 - 90 \rho ^2\right) \nonumber \\
	&~ &\, &-7 \hat{M}_X^6 - (11 \rho + 6) \hat{M}_X^4 - 45 \rho  \hat{M}_X^2  + 45 \rho ^2  \bigg)\delta^{(3)}(\hat{M}^2_X -\rho) \nonumber \\
	&~ &\, &-\frac{2}{15}\bigg(  108 \left(\hat{M}_X^2 - \rho \right) \hat{E}_0^4 - 30 \left(2 \hat{M}_X^4 - 3 \rho  \hat{M}_X^2 + \hat{M}_X^2 + (\rho - 1) \rho \right) \hat{E}_0^3 \nonumber \\
	&~ &\, &+ \left(-86 \hat{M}_X^4 + 16 \rho  \hat{M}_X^2 + 70 \rho ^2\right) \hat{E}_0^2 + 15 \left(3 \hat{M}_X^6 + (1 - 4 \rho ) \hat{M}_X^4 + \rho ^2 \hat{M}_X^2 - \rho ^2\right) \hat{E}_0 \nonumber \\
	&~ &\, &+ 8 \hat{M}_X^2 \left(\hat{M}_X^4 + 4 \rho  \hat{M}_X^2 - 5 \rho ^2\right)\bigg)\delta^{(4)}(\hat{M}^2_X -\rho)\bigg) \bigg\rbrace
\end{alignat}

\subsection{Differential rate $\text{d} \Gamma/\text{d}y  $}
We use in this Differental rate the dimensionless variable
\begin{equation}
	y  = 2\frac{E_e}{m_b} 
\end{equation}
Furthermore x is a short hand notation for \( x = \frac{y}{1-y} (1-y-\rho) \).
$\delta^{(n)}(x)$ means therefore the nth derivative of the delta function with respect to its argument x.
To rewrite the argument and derivative of the delta function linear in y, we use
\[ \delta^{(n)} (g(y)) = \left(\frac{1}{g'(y)} \frac{\text{d}}{\text{d} y} \right)^n \sum_i \frac{1}{|g'(y_i)|} \delta(y-y_i)\]
where
\[\left(\frac{1}{g'(y)} \frac{\text{d}}{\text{d} y} \right)^n = \underbrace{\frac{1}{g'(y)} \frac{\text{d}}{\text{d} y} \ldots \frac{1}{g'(y)} \frac{\text{d}}{\text{d} y}}_{\text{n times}} \]
\begin{alignat}{2}
	\frac{\text{d} \Gamma}{\text{d} y} = &\frac{G_F^2 m_b^5}{192 \pi^3} |V_{cb}|^2&\, &\bigg\lbrace 2 y^2 \bigg(\frac{(y-3) \rho ^3}{(y-1)^3}-\frac{3 \rho ^2}{(y-1)^2}-3 \rho -2 y+3 \bigg) \nonumber \\
	 &- \frac{2 y^3}{3 m_b^2} \hat{\mu}_\pi^2 &\, &\bigg( \frac{2 \left(y^2-5 y+10\right) \rho ^3}{(y-1)^5}+\frac{(15-6 y) \rho ^2}{(y-1)^4}+5\bigg) \nonumber \\
	&+ \frac{2 y^2}{3 m_b^2} \hat{\mu}_G^2 &\, &\bigg( \frac{5 \left(y^2-4 y+6\right) \rho ^3}{(y-1)^4}-\frac{9 (y-2) \rho ^2}{(y-1)^3}+\frac{6 (2 y-3) \rho }{(y-1)^2}+5 y+6 \bigg) \nonumber \\
	&+ \frac{2}{3 m_b^3}\hat{\rho}_D^3 &\, &\bigg(\frac{y^2 \left(5 y^4-30 y^3+75 y^2-84 y+54\right) \rho ^3}{(y-1)^6} \nonumber \\
	&~ &\, &+\frac{y^2 \left(-5 y^3+25 y^2-50 y+42\right) \rho ^2}{(y-1)^5} +\frac{2 y^2 \left(2 y^2-3 y-3\right) \rho}{(y-1)^3}   \nonumber \\
	&~ &\, &+ \frac{2 y^2 \left(2 y^2-5 y+9\right)}{y-1}-\frac{(\rho +1) (\rho +2) (\rho -1)^4}{\rho ^2}  \delta (y+\rho -1)  \nonumber \\
	&~ &\, &+\bigg(\frac{\rho ^4\left(-y^2+3 y+\rho -2\right)}{(y-1)^8}-\frac{\rho^3 y}{(y-1)^5}+\frac{(y-2) \rho^2}{(y-1)^3}+\frac{\rho}{y-1}\bigg)y^4 \delta '(x)\bigg) \nonumber \\
	&+ \frac{2 y^2}{3 m_b^3} \hat{\rho}_{LS}^3 &\, &\bigg(\frac{\left(y^3-5 y^2+10 y+6\right) \rho ^3}{(y-1)^5}+\frac{3 \left(y^2-4 y+6\right) \rho ^2}{(y-1)^4}+\frac{6 (2 y-3) \rho }{(y-1)^2}+4y +6\bigg) \nonumber \\
	&+ \frac{1}{15 m_b^4} s_1 &\, &\bigg(-\frac{6 y^2 \left(17 y^5-119 y^4+357 y^3-595 y^2+520 y-240\right) \rho ^3}{(y-1)^7} \nonumber \\
	&~ &\, &+\frac{2 y^2 \left(43 y^4-258 y^3+645 y^2-760 y+420\right) \rho ^2}{(y-1)^6} \nonumber \\
	&~ &\, &+\frac{20 y^2 \left(4 y^3-15 y^2+20 y-12\right) \rho }{(y-1)^4} +\frac{40 y^2 \left(2 y^2-4 y+3\right)}{(y-1)^2}\nonumber \\
	&~ &\, &\frac{2 \left(3 \rho ^4+10 \rho ^3-14 \rho ^2-34 \rho -45\right)  (\rho -1)^3}{\rho ^3}\delta (y+\rho -1) \nonumber 
\end{alignat}

\begin{alignat}{2}
	&~ &\, &+\bigg(-\frac{3 (y-19) \rho ^5 }{(y-1)^9} +\frac{\left(-2 y^2-28 y+75\right) \rho ^4 }{(y-1)^8}-\frac{2 (5 y+8) \rho ^3 y}{(y-1)^6}\nonumber \\
	&~ &\, &-\frac{2 (11 y+2) \rho ^2 y}{(y-1)^4}-\frac{3\left(y^2+4 y+5\right) \rho}{(y-1)^2}-3\bigg)y^4 \delta '(x) \nonumber \\
&~ &\, &+\bigg(+\frac{3 \rho ^6 }{(y-1)^{10}}+\frac{(5-2 y) \rho^5 }{(y-1)^9}-\frac{2 \rho ^4 y}{(y-1)^7} \nonumber \\
	&~ &\, &-\frac{2 \rho ^3 y}{(y-1)^5}+\frac{(3 y-5) \rho ^2}{(y-1)^3}+\frac{3 \rho }{y-1}\bigg) y^5 \delta ''(x) \bigg) \nonumber\\
	&+ \frac{1}{60 m_b^4} s_2 &\, &\bigg(\frac{6 y^2 \left(27 y^5-189 y^4+567 y^3-945 y^2+820 y-340\right) \rho ^3}{(y-1)^7} \nonumber \\
	&~ &\, &-\frac{6 y^2 \left(21 y^4-126 y^3+315 y^2-400 y+220\right) \rho ^2}{(y-1)^6}+\frac{40 y^3 \left(5 y^2-16   y+14\right) \rho }{(y-1)^4} \nonumber \\
	&~ &\, &+\frac{40 (y-2) y^3}{(y-1)^2}-\frac{2 \left(3 \rho ^3+23 \rho ^2-11 \rho +45\right)  (\rho -1)^4}{\rho ^3}\delta (y+\rho -1) \nonumber \\
	&~ &\, &+\bigg(\frac{3 (y-19)\rho ^5}{(y-1)^9}+\frac{\left(12 y^2+128 y-185\right) \rho^4}{(y-1)^8}-\frac{2  \left(15 y^2+32 y-105\right) \rho ^3}{(y-1)^6} \nonumber \\
	&~ &\, &+\frac{2\left(6 y^2-8 y-45\right) \rho ^2}{(y-1)^4}+\frac{\left(3 y^2+12 y+5\right) \rho }{(y-1)^2}+3 \bigg)y^4 \delta'(x) \nonumber \\
	&~ &\, &+\bigg(-\frac{3 \rho^6}{(y-1)^{10}}+\frac{3 (4 y-5) \rho ^5}{(y-1)^9}+\frac{6 (5-3 y) \rho ^4}{(y-1)^7} \nonumber \\
	&~ &\, &+\frac{6 (2 y-5) \rho   ^3}{(y-1)^5}-\frac{3 (y-5) \rho ^2}{(y-1)^3}-\frac{3 \rho }{y-1}\bigg) y^5 \delta ''(x)\bigg) \nonumber \\
	&+ \frac{1}{60 m_b^4}  s_3 &\, &\bigg(\frac{2 y^2 \left(-29 y^5+203 y^4-609 y^3+1015 y^2-1240 y+480\right) \rho^3}{(y-1)^7} \nonumber \\
	&~ &\, &+\frac{2 y^2 \left(7 y^4-42 y^3+105 y^2-400 y+240\right) \rho ^2}{(y-1)^6} \nonumber \\
	&~ &\, &-\frac{40 y^2 \left(3 y^3-11 y^2+14 y-3\right) \rho }{(y-1)^4}-\frac{40 y^2 \left(4 y^2-14 y+9\right)}{(y-1)^2} \nonumber \\
	&~ &\, &+\frac{2 \left(17 \rho ^2+34 \rho +45\right) (\rho -1)^5}{\rho ^3} \delta (y+\rho -1) \nonumber \\
	&~ &\, &+\bigg(-\frac{ (37 y+17) \rho ^5}{(y-1)^9}+\frac{ \left(52 y^2-32 y-65\right) \rho^4}{(y-1)^8}+\frac{2 \left(5 y^2-12 y+45\right) \rho ^3}{(y-1)^6} \nonumber \\
	&~ &\, &-\frac{2 \left(14 y^2-32 y+25\right) \rho^2}{(y-1)^4}+\frac{\left(3 y^2-28 y+5\right) \rho }{(y-1)^2}+3 \bigg) y^4 \delta '(x) \nonumber \\
	&~ &\, &+\bigg(-\frac{3 \rho^6}{(y-1)^{10}}+\frac{3 (4 y-5) \rho ^5}{(y-1)^9}+\frac{6 (5-3 y) \rho ^4}{(y-1)^7} \nonumber \\
	&~ &\, &+\frac{6 (2 y-5) \rho ^3}{(y-1)^5}-\frac{3 (y-5)\rho ^2}{(y-1)^3}-\frac{3 \rho }{y-1}\bigg)y^5 \delta ''(x) \bigg) \nonumber 
\end{alignat}

\begin{alignat}{2}
	&+ \frac{1}{10 m_b^4} s_4 &\, &\bigg(\frac{2 y^2 \left(y^4-6 y^3+15 y^2-20 y-10\right) \rho ^3}{(y-1)^6}-\frac{y^2 \left(7 y^3-35 y^2+55 y+45\right) \rho ^2}{(y-1)^5} \nonumber \\
	&~ &\, &-\frac{2 y^2 \left(2 y^3-5 y^2+15\right) \rho }{(y-1)^4} +5 y^2 +\frac{2 (\rho -2) (\rho +1)  (\rho -1)^4}{\rho ^2}\delta (y+\rho -1)\nonumber \\
	&~ &\, &+\bigg(-\frac{2\rho ^5}{(y-1)^8}+\frac{6 (y-2)\rho ^4}{(y-1)^7}-\frac{2 \left(3 y^2-13 y+12\right) \rho ^3}{(y-1)^6} \nonumber \\
	&~ &\, &+\frac{2 \left(y^2-7 y+10\right) \rho^2}{(y-1)^4}+\frac{2 (y-3) \rho }{(y-1)^2}\bigg)y^4 \delta '(x) \bigg) \nonumber \\
	&+ \frac{1}{30 m_b^4} s_5 &\, &\bigg(\frac{4 y^2 \left(19 y^4-114 y^3+285 y^2-380 y+60\right) \rho^3}{(y-1)^6} \nonumber \\
	&~ &\, &+\frac{y^2\left(-11 y^3+55 y^2-65 y+285\right) \rho ^2}{(y-1)^5}+\frac{2 y^2 \left(74 y^3-215 y^2+120 y+165\right)   \rho }{(y-1)^4} \nonumber \\
	&~ &\, &+\frac{15 (19-11 y) y^2}{y-1}+\frac{2 \left(13 \rho ^2-3 \rho +74\right)  (\rho -1)^4}{\rho ^2}\delta (y+\rho -1) \nonumber \\
	&~ &\, &+\bigg(-\frac{26 \rho ^5}{(y-1)^8}+\frac{2 (19 y-8) \rho ^4}{(y-1)^7}+\frac{2 \left(y^2-31y+54\right) \rho ^3}{(y-1)^6} \nonumber \\
	&~ &\, &-\frac{2 \left(7 y^2-39 y+80\right) \rho ^2}{(y-1)^4}+\frac{2 (31-7 y) \rho }{(y-1)^2}\bigg) y^4 \delta '(x)\bigg) \bigg\rbrace
\end{alignat}

\subsection{Total decay rate $\Gamma$}
\begin{alignat}{2}
	\Gamma = &\frac{G_F^2 m_b^5}{192 \pi^3} |V_{cb}|^2&\,&\bigg\lbrace -\rho ^4+8 \rho ^3-12 \log (\rho ) \rho ^2-8 \rho +1 \nonumber \\
	&-\frac{1}{2 m_b^2} \hat{\mu}_\pi^2 &\,&\bigg( -\rho ^4+8 \rho ^3-12 \log (\rho ) \rho ^2-8 \rho +1\bigg) \nonumber \\
	&+ \frac{1}{2 m_b^2} \hat{\mu}_G^2 &\,&\bigg( -5 \rho ^4+24 \rho ^3-12 (\log (\rho )+2) \rho ^2+8 \rho -3 \bigg) \nonumber \\
	&+ \frac{2}{3 m_b^3} \hat{\rho}_D^3 &\,&\bigg( -5 \rho ^4+16 \rho ^3-12 \rho ^2-16 \rho +12 \log (\rho )+17 \bigg) \nonumber \\
	&+ \frac{8}{9 m_b^4} s_1 &\,&\bigg(9 \rho ^4-20 \rho ^3+9 \rho ^2+6 \log (\rho )+2 \bigg) \nonumber \\
	&+ \frac{1}{9 m_b^4} s_2 &\,&\bigg(-27 \rho ^4+76 \rho ^3-72 \rho ^2+36 \rho -12 \log (\rho )-13 \bigg) \nonumber \\
	&+ \frac{4}{9 m_b^4} s_3 &\,&\bigg( 3 \rho ^4-7 \rho ^3+9 \rho ^2-21 \rho +4 (3 \log (\rho )+4) \bigg) \nonumber \\
	&+ \frac{1}{3 m_b^4} s_5 &\,&\bigg(-5 \rho ^4+16 \rho ^3-12 \rho ^2-16 \rho +12 \log (\rho )+17 \bigg) \bigg\rbrace
\end{alignat}

\end{document}